\documentclass[english,showpacs, twocolumn,amsmath,amssymb,showkeys]{revtex4-1}

\newcommand{\LAO}{LaAlO$_3$ }
\newcommand{\STO}{SrTiO$_3$ }

\newcommand{\LAOSTO}{LaAlO$_3$/SrTiO$_3$ }

\usepackage[T1]{fontenc}
\usepackage[latin9]{inputenc}
\usepackage{color}
\usepackage{units}
\usepackage{amssymb}
\usepackage{graphicx}
\usepackage{esint}
\usepackage{bm}
\usepackage{natbib}
\usepackage{ulem}
\usepackage{multirow}
\usepackage{mathrsfs}
\usepackage{braket}
\usepackage[colorlinks=true, citecolor=red]{hyperref}
\setcitestyle{journalcolor= blue}

\begin{document}

\title{Observation of transient superconductivity at \LAOSTO interface}

\author{Gopi Nath Daptary}
\affiliation{Department of Physics, Indian Institute of Science, Bangalore 560012, India}
\author{Shelender Kumar}
\affiliation{Department of Physics, Indian Institute of Science, Bangalore 560012, India}
\author{Pramod Kumar}
\affiliation{National Physical Laboratory, New Delhi 110012, India}
\author{Anjana Dogra}
\affiliation{National Physical Laboratory, New Delhi 110012, India}
\author{Dushyant Kumar}
\affiliation{Condensed Matter-Low Dimensional Systems Laboratory, Department of Physics, Indian Institute of Technology Kanpur, Kanpur 208016, India}
\author{N. Mohanta}
\affiliation{Center for Electronic Correlations and Magnetism, 
	Theoretical Physics III, Institute of Physics
	University of Augsburg, 86135 Augsburg, Germany}
\author{A. Taraphder}

\affiliation{Department of Physics, Indian Institute of Technology Kharagpur, W.B. 721302, India}
\author{R. C. Budhani}
\affiliation{National Physical Laboratory, New Delhi 110012, India}
\author{Aveek Bid}
\email{aveek.bid@physics.iisc.ernet.in}
\affiliation{Department of Physics, Indian Institute of Science, Bangalore 560012, India.}

\begin{abstract}
We report the observation of a magnetic field assisted transient superconducting state in the two dimensional electron gas existing at the interface of \LAOSTO heterostructures. This metastable state depends critically on the density of charge carriers in the system. It appears concomitantly with a Lifshitz transition as a consequence of the interplay between ferromagnetism and superconductivity and the finite relaxation time of the in-plane magnetization. Our results clearly demonstrate the inherently metastable nature of the superconducting state competing with a magnetic order in these systems. The co-existence of superconductivity and ferromagnetism in the conducting electronic layer formed at the interface of insulating oxides has thrown up several intriguing and as yet unanswered questions. An open question in this field is the energetics of the interplay between these two competing orders and the present observation goes a long way in understanding the underlying mechanism. 
\end{abstract}

\maketitle
%
%

\section{Introduction}
The mutual interplay of point group symmetry, charge inversion symmetry, U(1) gauge symmetry and  spin rotation symmetry in heterostructures of complex perovskite oxides~\cite{ohtomo2004high} lead to the co-existence of a host of intriguing properties - ferroelasticity, ferroelectricity, superconductivity and ferromagnetism~\cite{doi:10.1146/annurev-matsci-070813-113437,hwang2012emergent}. Superconductivity and magnetism are generally considered to be incompatible with each other and hence reports of the observation of a possible co-existence of these two phases in the conducting electronic layer formed at the interface of two insulating oxides \LAO and \STO~\cite{ohtomo2004high, doi:10.1146/annurev-matsci-070813-113437,hwang2012emergent, reyren2007superconducting, mannhart2008two,zubko2011interface} has opened up a new direction of research in condensed matter physics. In this paper, we report the observation of magnetic-field assisted transient superconducting state (TSS) at the interface of \LAO and \STO at 245 mK. The TSS appears concomitantly with a Lifshitz transition in the system as a consequence of the interplay between ferromagnetism and superconductivity and the finite relaxation time of in-plane magnetization. To the best of our knowledge such a  transient superconducting state has not been observed in condensed matter systems.
Despite intensive research over the last decade~\cite{doi:10.1146/annurev-matsci-070813-113437} the co-existence of superconducting and ferromagnetic phases in this system is still debatable. There is now overwhelming evidence that superconductivity in \LAOSTO is mediated by phonons~\cite{phonon} and is conventional BCS-like~\cite{richter}. Scanning superconducting quantum interference device (SQUID) measurements have revealed that the superconductivity in these systems is probably spatially inhomogeneous~\cite{bert2011direct} although more recent experiments may suggest otherwise~\cite{PhysRevB.85.224518}. Direct measurements of the magnetization in this system have yielded contrasting results. On one hand torque measurements show a large in-field  magnetization of 0.3-0.4 $\mu_B$ per interfacial Ti ion~\cite{li2011coexistence}. On the other hand scanning SQUID  experiments show that there are only spatially inhomogeneous  patches of local moments with no net magnetization~\cite{bert2011direct}. Although various scenarios have been invoked to reconcile these apparently contradictory experimental observations~\cite{michaeli2012superconducting,mohanta2014phase,banerjee2013ferromagnetic,PhysRevB.82.165127,coey2016surface} a clear picture of the magnetization behaviour of this system is yet to emerge.

\begin{figure}[tbh]
	\begin{center}
		\includegraphics[width=0.48 \textwidth]{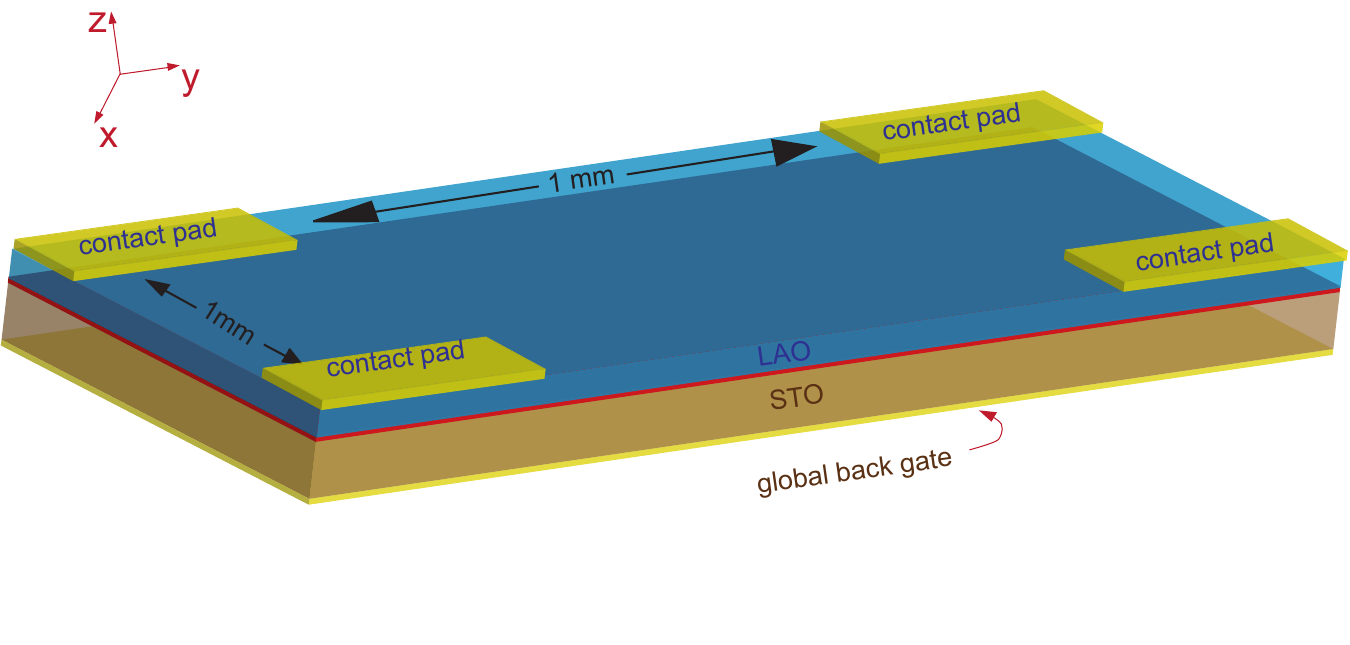}
		\small{\caption{ A schematic of the device structure - the red shaded area represents the 2DEG located at the interface of \LAO and \STO. \label{fig:figureS1}}}
	\end{center}
\end{figure}

When the thickness of the \LAO layer grown over TiO$_{2}$ terminated \STO layer exceeds 4 unit cells, 0.5 electrons per unit cell are transferred from the top layer of \LAO to the Ti$^{3+}$ ions at the interface to avoid a polar catastrophe~\cite{mannhart2008two} resulting in a quasi 2-dimensional electron gas (2DEG) at the interface. Hall measurements suggest that a very small fraction of these electrons actually take part in transport - it is believed that most of them get localized in the d-t$_{2g}$ orbitals of the Ti atoms at the interface because of strong on site Hubbard and nearest neighbor Coulomb repulsive interactions forming the local moments responsible for ferromagnetism~\cite{pavlenko2012magnetic}. The breaking of mirror inversion symmetry at the interface lifts the degeneracy of the t$_{2g}$ levels of Ti ions at the interface~\cite{khalsa2013theory}  with the d$_{xy}$ level having a lower energy than the d$_{xz}$ and d$_{yz}$ orbitals~\cite{salluzzo2009orbital}. At low number densities all the conduction electrons occupy the lower lying d$_{xy}$ orbitals at the interface~\cite{doi:10.1146/annurev-matsci-070813-113437}. It has recently been proposed that when the number density $n_s$ of itinerant electrons exceeds a certain critical value the system undergoes a Lifshitz transition at which point the $d_{xz}/d_{yz}$ bands near the interface begin to get occupied. The system now effectively has two types of carriers - a high density electron gas residing in the $d_{xy}$ orbital and a lower density high mobility electron gas occupying the $d_{xz}/d_{yz}$ orbitals~\cite{aplmultiple,biscaras2012two,biscaras2010two,kim2010nonlinear}. An additional parameter controlling this system is the strong Rashba spin orbit coupling (SOC) arising due to the broken inversion symmetry at the interface which allows the electronic properties of the system to be modulated over a large range by means of a gate voltage induced electric field ~\cite{caviglia2010tunable}. Oxygen vacancies, controlled by the O$_2$ partial pressure during the deposition of \STO, are also believed to play a crucial role in determining the magnetic and electrical transport properties of this system~\cite{kalabukhov2007effect, wang2011electronic}.  

 \begin{figure}[t]
 	\begin{center}
 		\includegraphics[width=0.48\textwidth]{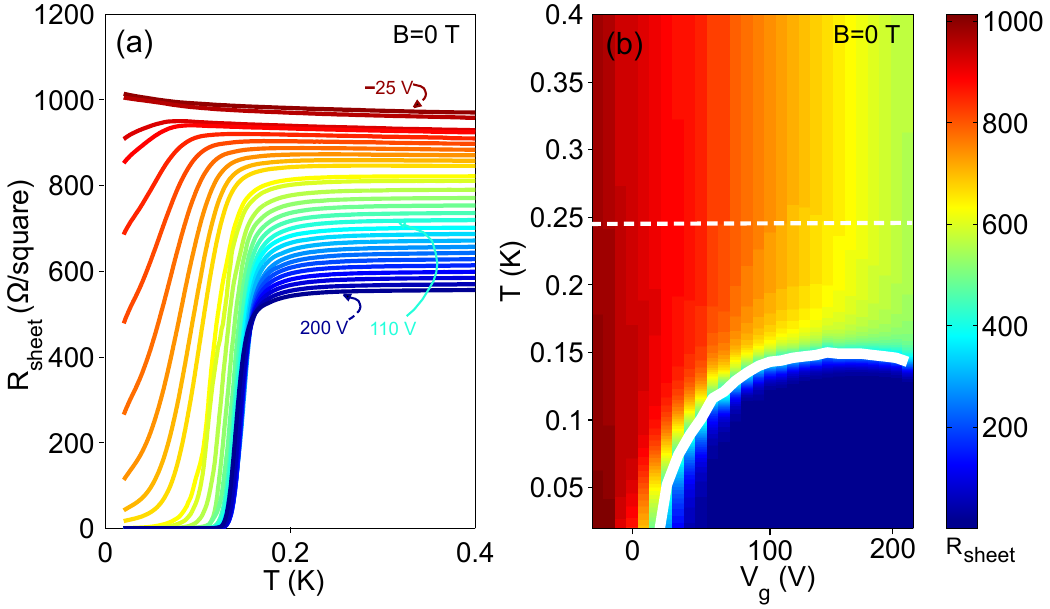}
 		\small{\caption{ (a) Sheet resistance $R_{sheet}$ of the device S2 measured as a function of temperature at different values of $V_g$ ranging from -25~V till 200~V (in steps of 5~V from -25~V to 0~V and subsequently in steps of 10~V from 0~V to 200~V). (b) Resistance $R_{sheet}$ in color scale as a function of temperature $T$ and gate voltage $V_g$. The white solid line superimposed on the plot shows the superconducting transition temperature. \label{fig:figure1}}}
 	\end{center}
 \end{figure}

\section{Results}
\subsection{Sample preperation}

Our measurements were performed on samples with 10 unit cells of LaAlO$_{3}$ grown by Pulsed Laser Deposition (PLD) on TiO$_{2}$ terminated (001) SrTiO$_{3}$ single crystal substrates of thickness 0.5~mm~\cite{0953-8984-27-12-125007}. As received SrTiO$_{3}$ substrates were pre-treated with standard buffer HF solution~\cite{kawasaki1994atomic} in order to achieve uniform TiO$_{2}$ termination which was confirmed from atomic force microscopy measurements. Prior to deposition, the treated substrates were annealed for an hour at 830 ${}^\circ$C in oxygen partial pressure of $7.4\times10^{-2}$ mbar to remove any moisture and organic contaminants from the surface and also to reconstruct the surface so that pure TiO$_2$ termination was realized. This was followed by the deposition of  10 unit cells LaAlO$_{3}$ at 800 ${}^\circ$C at an oxygen partial pressure of $1\times10^{-4}$ mbar.  Growth with the precision of single unit cell was monitored by the oscillations count using in-situ RHEED gun. Post-deposition, the samples were cooled at the same $O_2$ partial pressure at the rate of 10 ${}^\circ$C/min to the ambient temperature. The epitaxial nature of the films was confirmed by HRXRD performed on a 20 u.c. \LAO film grown under identical conditions on TiO$_2$ terminated \STO which allowed us to measure the c-axis lattice parameter of LaAlO$_3$. The thickness of one unit cell from these measurements came out to be 3.75~\AA~\cite{0953-8984-27-12-125007}.

A schematic of the device structure is shown in figure~\ref{fig:figureS1}. Electrical contacts were created on top of the \LAO substrate by thermal evaporation of 5~nm Cr followed by 100~nm of Au and  were wire bonded to the measurement chip carrier. A gold film deposited on the back side of the \STO substrate acted as one plate of the capacitor while the conducting layer acted as the other plate of the capacitor for electrostatic gating of the device. The \STO substrate  acted as the gate dielectric material. Measurements were performed on five different samples grown under similar conditions, they differed only in their carrier concentrations at zero gate voltage. All the devices showed qualitatively the same behavior. In this paper we present the results of detailed measurements on two devices - S2 and S5 with S2 having a slightly lower sheet number density of charge carriers ($n_s \approx 1.65\times 10^{13}$ cm$^{-2}$ at 250 mK) as compared to S5 ($n_s \approx 2\times 10^{13}$ cm$^{-2}$ at 250mK).  The measurements were performed down to 245~mK in a He-3 refrigerator and down to 10~mK in a dilution refrigerator.

 \begin{figure}[t]
 	\begin{center}
 		\includegraphics[width=0.48\textwidth]{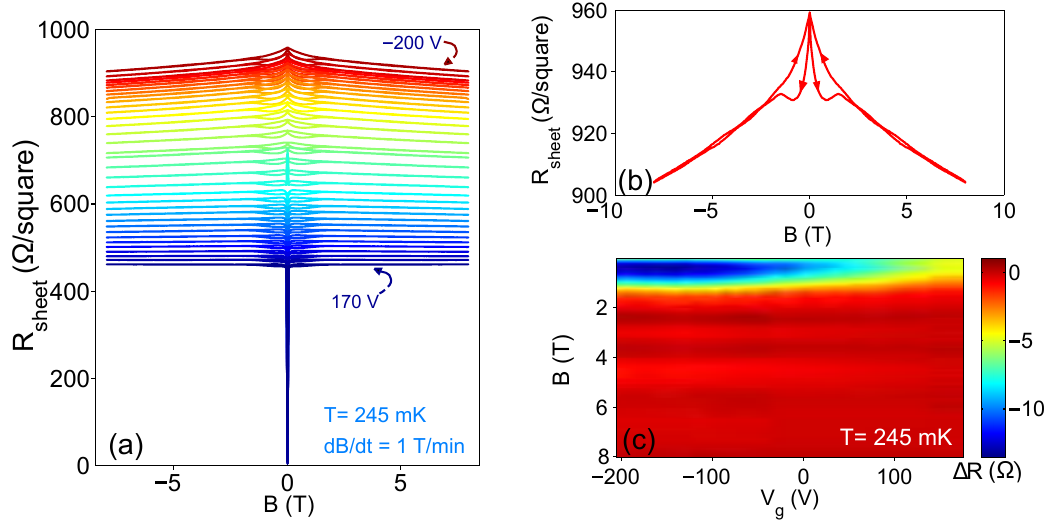}
 		\small{\caption{(a) Magnetoresistance of the device S2 at different gate voltages ranging from -200~V till 170~V (in steps of 20~V from -200~V till -150~V and subsequently in steps of 10~V from -150~V till 170~V) measured at 245~mK. The TSS state appears for values of gate voltages $V_g>V_g^*$. (b) Magnetoresistance at gate voltage V$_{g}$ = -200~V showing hysteresis at low magnetic fields. The arrows denote the direction of magnetic field sweep.(c) Hysteresis in magnetoresistance as a function of gate voltage and magnetic field at 245~mK. Note that the hysteresis gradually disappears with increasing $V_g$. \label{fig:figure2}}}
 	\end{center}
 \end{figure}

\begin{figure*}[t]
	\begin{center}
		\includegraphics[width=0.8\textwidth]{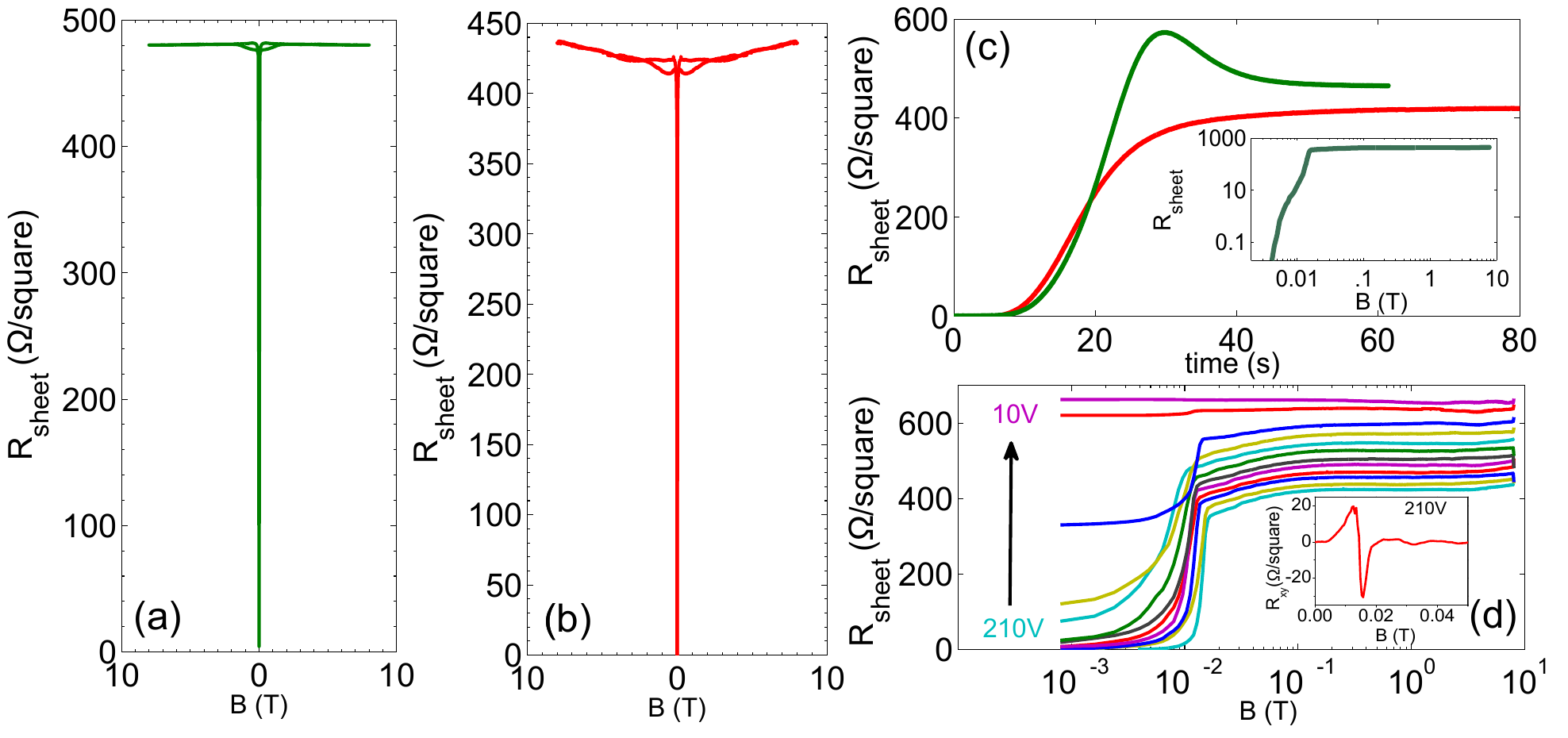}
		\small{{\caption{(a) Magnetoresistance of the device S2 measured at gate voltage $V_{g}$ = 150~V and temperature 245 mK showing the appearance of the TSS. (b)Magnetoresistance of the device S5 measured at gate voltage $V_{g}$ = 210~V and temperature 245 mK showing the appearance of the  TSS. In both (a) and (b) the magnetic field was swept down from $B=8$~T at a rate $dB/dt = 1$~ T/min. (c) Time relaxation of the TSS for the two devices (green curve: device S2, $V_g$=135~V and  red curve: device S5 and $V_g$=210~V) measured after the magnetic field was swept down to 0~T at a  rate $dB/dt$ = 1~T/min. The measurements were performed at  245~mK. Inset: log-log plot of the same data as plotted in (b) to show the detailed evolution of the TSS with magnetic field. (d)  Magnetoresistance of device S5 at different values of $V_g$ at 245~mK - the data was acquired as  the B field was swept down from 8~T at a rate $dB/dt = 1$~ T/min. The values of $V_g$ range from 210~V to -10~V in steps of 20~V. Inset: Hall data obtained close to B=0~T at 210~V as the magnetic field was swept down to 0~T at a  rate $dB/dt$ = 1~T/min.
					\label{fig:figure3}}}}
	\end{center}
\end{figure*}

 \begin{figure}[t]
 	\begin{center}
 		\includegraphics[width=0.48\textwidth]{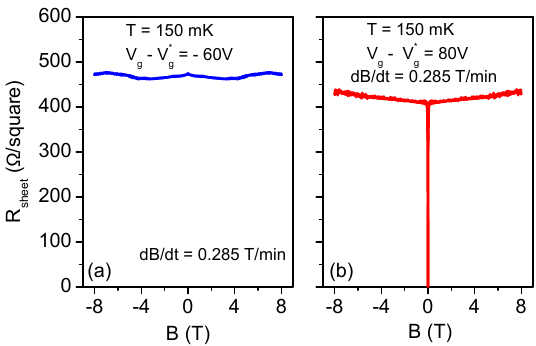} \small{\caption{ Magnetoresistance measured at two values of $V_g$ around the doping level corresponding to $V_g^*$. The measurement were done at 150 mK, the $B$ field sweep rate was 0.285 T/min.\label{fig:figureS5}}}
 	\end{center}
 \end{figure}

\begin{figure}[t]
	\begin{center}
		\includegraphics[width=0.48\textwidth]{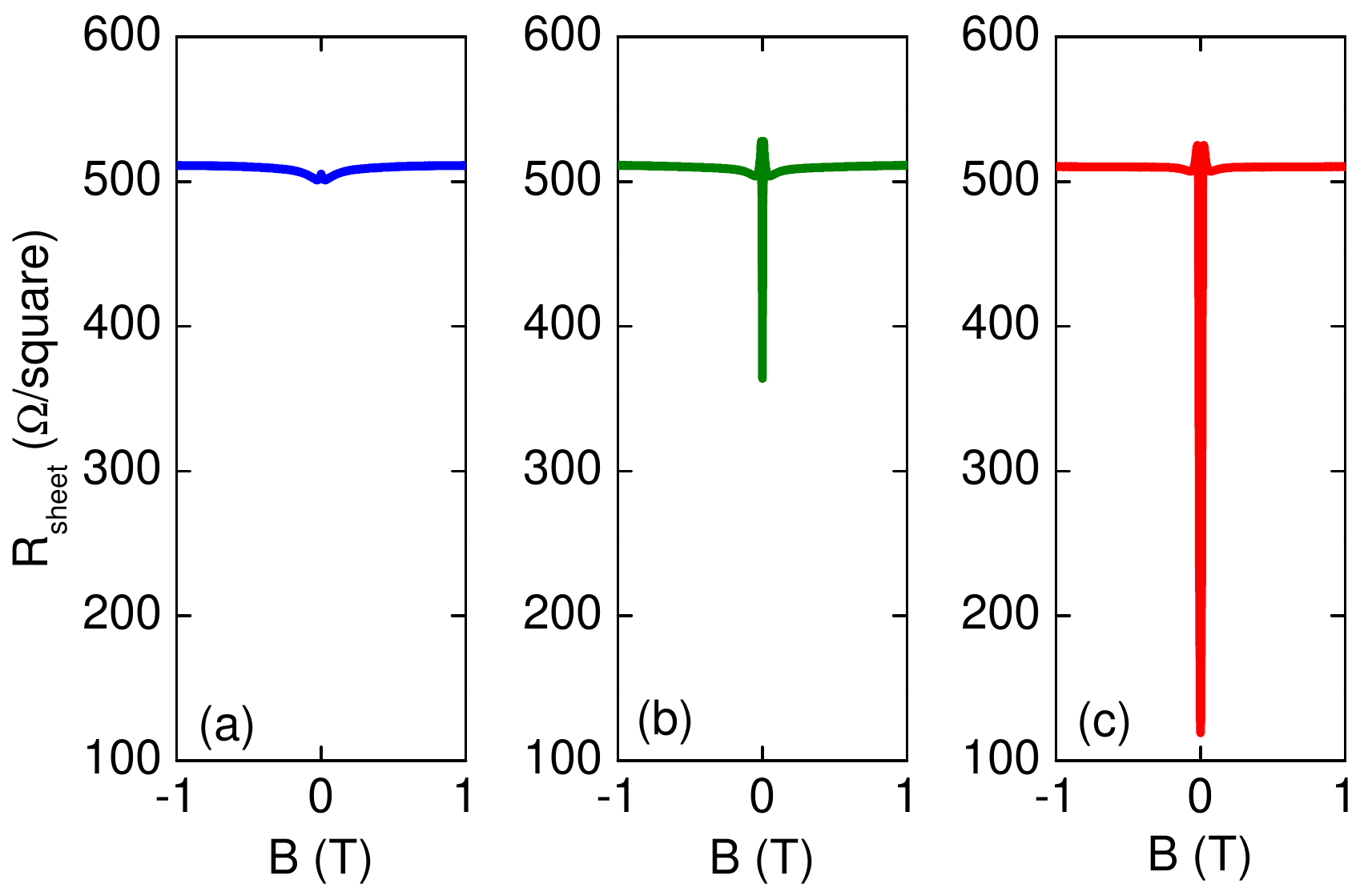}
		\small{\caption{ Effect of different sweep rates of the magnetic field on the transient superconducting state  - (a) $dB/dt=0.1$ T/min, (b) $dB/dt=0.2$ T/min and (c) $dB/dt = 0.5$ T/min. The measurements were done at $V_g = 110$ V and device temperature 245 mK. \label{fig:figureS4}}}
	\end{center}
\end{figure}

\subsection{Resistance and Magnetoresistance}
The sheet resistance $R_{sheet}$ of the device  S2 as a function of temperature at different gate voltages $V_{g}$ is shown in Figure \ref{fig:figure1}(a). The superconducting transition temperature $T_C$ and the normal state resistance were both found to depend sensitively on the gate voltage. The $T_C$ (defined as the temperature where resistance drops to 50\% of its normal state resistance) increases as the system is progressively electron doped (see Figure \ref{fig:figure1}(b)) in conformity with previous observations in similar systems~\cite{caviglia2008electric, dikin2011coexistence, biscaras2010two}.

The magnetoresistance data for magnetic fields applied perpendicular to the interface measured at a few representative values of $V_{g}$ are shown in Figure \ref{fig:figure2}(a) for the device S2. The measurements were performed at 245~mK where the device is in the normal state at all measured gate voltages (shown by the dotted white line in figure~\ref{fig:figure1}(b)). We notice a distinct change in the nature of the magnetoresistance curves as $V_g$ changes from a large negative value to a large positive value, the change occurring around a critical doping level $n^*$ corresponding to a gate voltage $V_g$ = $V_g^*$. The value of $V_g^*$ is sample specific, depending on the initial doping level of the device, for this particular device $V_g^*$ = 110 V. Later in this article we discuss the physical significance of $n^*$. In the low carrier doping regime ($n<n^*$), the magnetoresistance is negative, quite small in magnitude (about $4\%$ at 8 T field) and is hysteretic (Figure \ref{fig:figure2}(b)). The hysteresis is time dependent and relaxes exponentially to an equilibrium value over a time scale of a few hundreds of seconds. With increasing $V_g$ both the magnitude of hysteresis and the relaxation time decreases and eventually vanishes at around the critical gate voltage $V_g$=110~V (see Figure \ref{fig:figure2}(c)). Although hysteresis in magnetoresistance in the low doping regime has been seen previously in \LAOSTO heterostructure devices and was taken to indicate the presence of ferromagnetic domains in the system~\cite{brinkman2007magnetic,dikin2011coexistence}, there is a growing concern in the community that it might also have contributions from induction effects due to fast magnetic field sweeps. We do not discuss further the data in this region of doping leaving it for further experimental analysis. 

\begin{figure}[t]
	\begin{center}
		\includegraphics[width=0.48\textwidth]{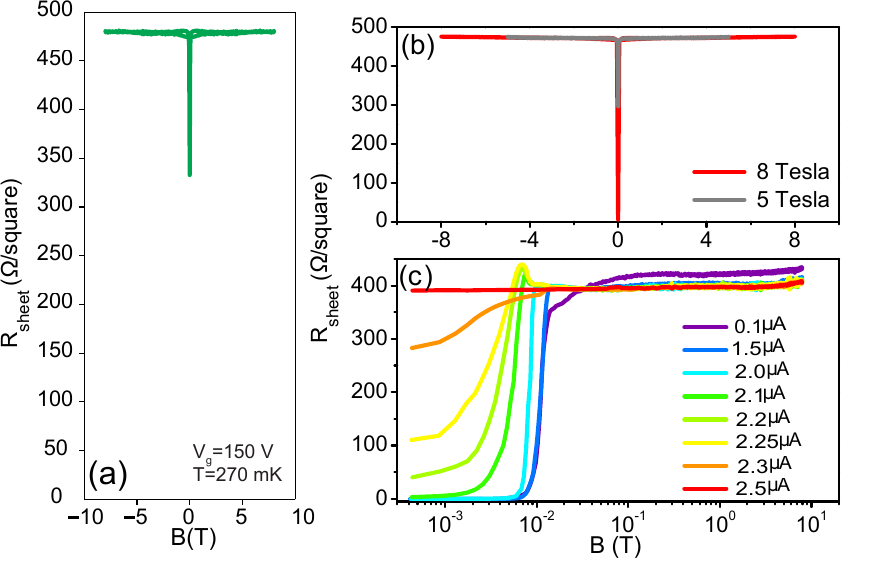}
		\small{{\caption{ (a) Magnetoresistance of device S2 at gate voltage $V_{g}$ = 150~V and temperature 270 mK  - $R_{sheet}$ shows a dip near $B$ = 0~T as $B$ is swept down but the system does not go into the TSS. The magnetic field was swept down from $B=8$~T at a rate $dB/dt = 1$~ T/min. (b) Effect of the maximum field on the TSS, the red curve shows the data taken as the magnetic field is swept down from $B = 8$~T while the gray curve shows the data taken as the magnetic field is swept down from $B = 5$~T. Note that for $B_{max} = 5$~T the transient superconducting state does not appear. The measurements were done at $V_g = 135$~V and 245~mK. (c)  TSS as a function of dc bias current measured at 250~mK, $V_g$=210~V. \label{fig:figure4}}}}
	\end{center}
\end{figure}

 \begin{figure}[t]
 	\begin{center}
 		\includegraphics[width=0.45\textwidth]{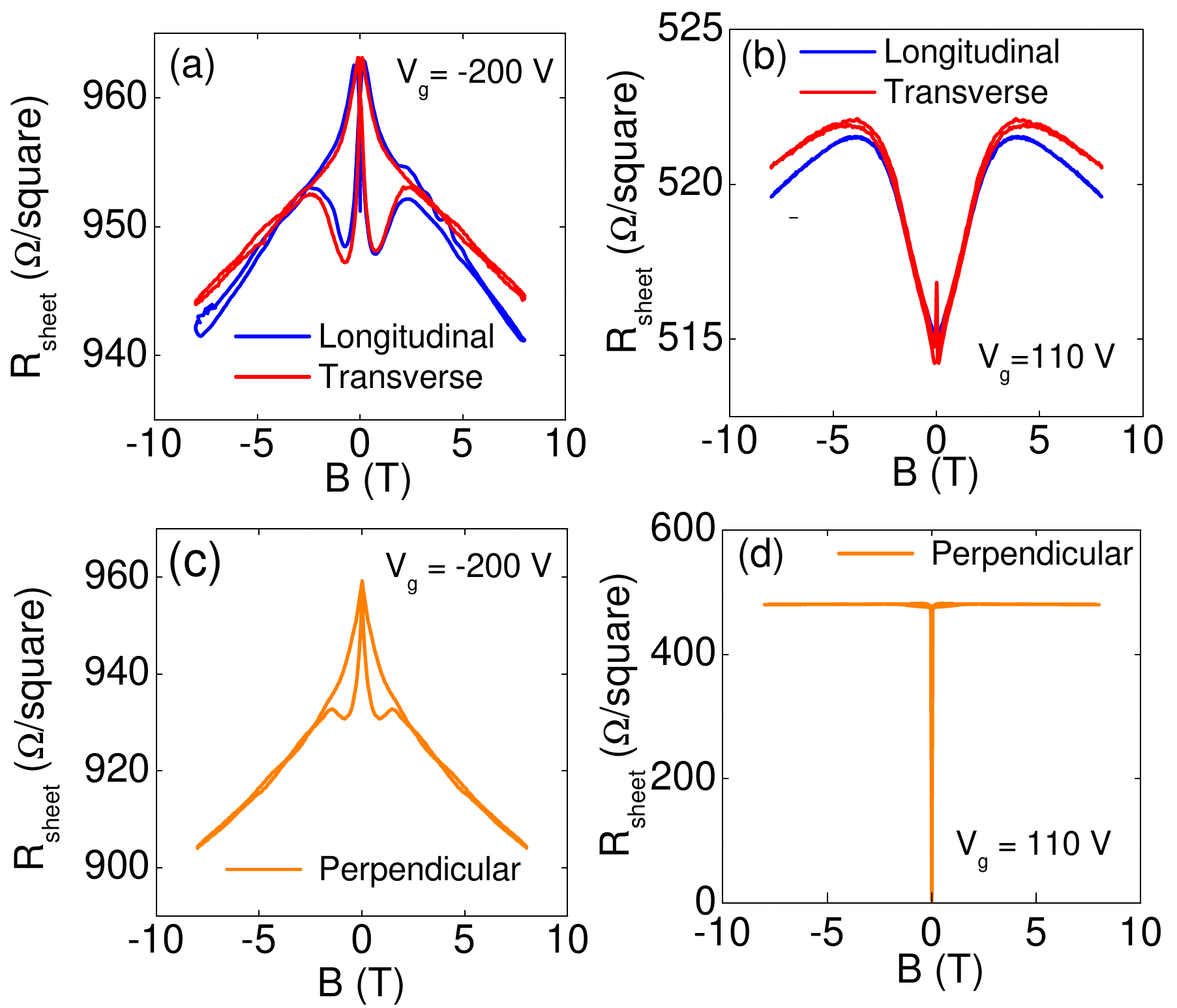}
 		\small{\caption{ Magnetoresistance of the device with the magnetic field applied parallel to the interface, the measurements were taken at 245 mK and (a) $V_g$ = -200~V, (b) $V_g$ = 110~V. In the longitudinal configuration the magnetic field was parallel to the conducting layer at the interface and also to the direction of the current. In the transverse configuration the magnetic field was parallel to the conducting layer at the interface and perpendicular to the direction of the current. For comparison we also plot the magnetoresistance with the magnetic field applied perpendicular to the interface, the measurements were taken at 245 mK and (c) $V_g$ = -200~V, (d) $V_g$ = 110~V. The TSS only appears for $V_g > V_g^*$ and only when the magnetic field direction is perpendicular to the conducting layer at the interface. In all cases, the $B$ field was swept at $dB/dt=1$~T/min.\label{fig:figureS10}}}
 	\end{center}
 \end{figure}

\begin{figure}[t]
	\begin{center}
		\includegraphics[width=0.45\textwidth]{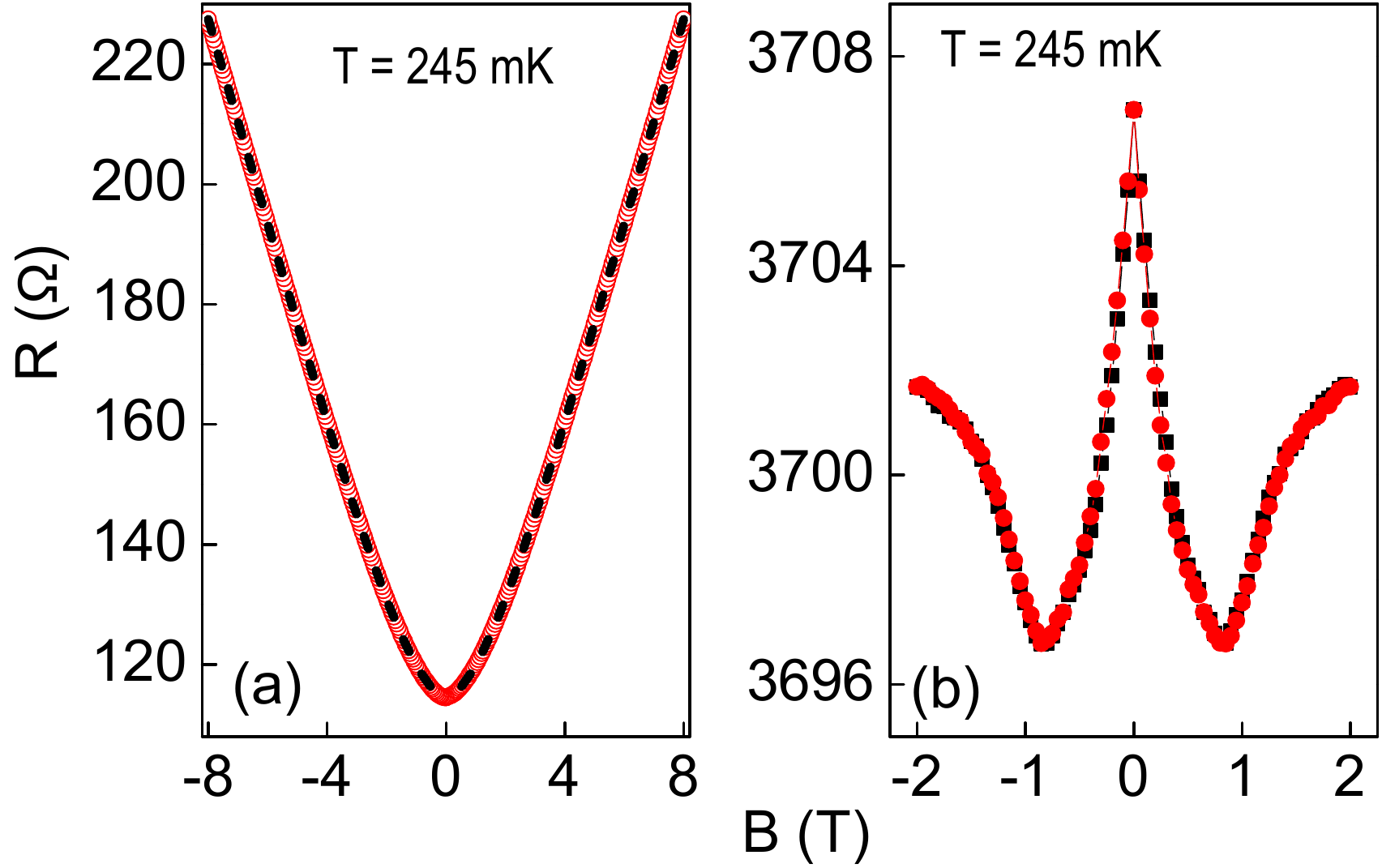}
		\small{\caption{(a) Plot of the magnetoresistance of reduced \STO measured with the magnetic field applied perpendicular to the plane of the device. Magnetoresistance  plots for increasing magnetic field (red open circles ) and for decreasing magnetic field (black dashed line) fall exactly on top of each other. (b) Plot of the magnetoresistance of LaTiO$_3$/SrTiO$_3$ heterostructure with the magnetic field applied perpendicular to the plane of the device. Magnetoresistance plots for increasing magnetic field (red circles) and for decreasing magnetic field (black squares) superimpose. In both cases there is no transient superconductivity or hysteresis in the magnetoresistance. Measurements were done at $T$ = 245~mK with $dB/dt=1$~T/min.\label{fig:figure7}}}
	\end{center}
\end{figure}

\begin{figure}[t]
	\begin{center}
		\includegraphics[width=0.48\textwidth]{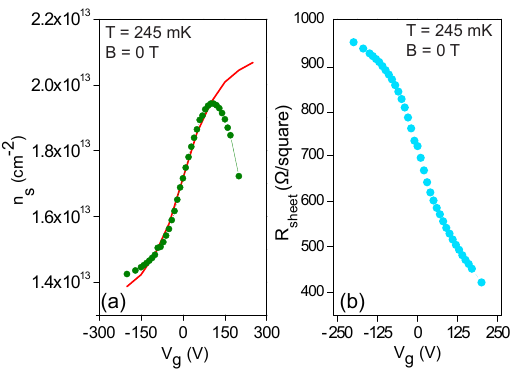}
		\small{{\caption{(a) Plot of $n_s$ (extracted  from Hall measurement data) as a function of $V_g$ (olive field circles) for device S2, the measurements were performed at 245 mK. The solid line is the expected value of $n_s(V_g)$ taking into account the electric field dependence of the dielectric constant of SrTiO$_{3}$. (b)  Zero field resistance $R_{sheet}$  of the device as a function of $V_g$. 
					\label{fig:figure5}}}}
	\end{center}
\end{figure}

\subsection{Magnetic field assisted transient superconductivity}

For $V_{g} > V_{g}^*$, the magnetoresistance is positive as the magnetic field is swept from 0 T to 8 T. This change from a positive magnetoresistance to a negative magnetoresistance around a certain value of $V_{g}$ has been observed before in \LAOSTO heterostructures and has been interpreted to be due to a transition from weak localization to weak anti-localization mediated by the large Rashba SOC present in this system~\cite{caviglia2010tunable}. As the magnetic field is swept back down towards 0 T, the magnetoresistance curve retraces itself till about 20 mT below which the sheet resistance jumps down by more than four orders of magnitude and the system goes superconducting with the resistance becoming smaller than our measurement resolution. The data from typical measurements are plotted in figures~\ref{fig:figure3}(a) and~\ref{fig:figure3}(b) for devices S2 and S5 respectively. In the inset of figure~\ref{fig:figure3}(c) we have re-plotted the data from device S5 in the low-field regime to emphasize the precipitous drop in the sheet resistance. [The corresponding Hall data is shown in the inset of figure~\ref{fig:figure3}(d)]. Note that in the absence of magnetic field, the devices are in non-superconducting state with $R \approx 425-475$~$\Omega$. The superconducting state thus reached is transient and relaxes back to the original zero field resistive state with a time constant of around 10~seconds (see Figure~\ref{fig:figure3}(c)). In figure~\ref{fig:figure3}(d) we show a plot of the magnetoresistance for device S5 at different values of $V_g$ - the data was acquired as  the B field was swept down from 8~T. It can be seen that the TSS state appears only for  values of gate voltage $V_g>V_g^*$.  This condition for the observation of the TSS held true even for temperatures quite close to the $T_C$ - the data obtained at 150~mK have been shown in figure~\ref{fig:figureS5}. However, due to technical limitations, the magnetic field sweep rate at these temperatures had to be limited to $dB/dt = 0.285$~ T/min. As shown in figure~\ref{fig:figureS4}, this sweep rate is not enough to take the resistance to zero. We however, see a large dip in the  resistance at zero field indicative of the TSS state only for $V_g>V_g^*$ - it can be seen that even at temperatures very close to $T_C$ ($T/T_C \sim 1.05$) we do not observe any signatures of TSS for $V_g<V_g^*$.

The appearance of this TSS depended critically on $dB/dt$, the rate at which the magnetic field was swept down from its maximum value. For slow sweep rates of the magnetic field, there appeared a dip in the resistance near 0 T, but the resistance remained finite (see figure \ref{fig:figureS4}). The magnitude of the dip increased as $dB/dt$ increased and beyond a certain value of $dB/dt$, the system went into the transient superconducting state.

  The TSS was observed upto about 260 mK. Beyond this temperature the TSS does not appear although a dip in the magnetoresistance is seen near zero magnetic field as the field is swept down from 8 T with the magnitude of the dip rapidly decreasing with increasing temperature. The data from a typical measurement at 270 mK and $V_g$ = 150 V are shown in Figure \ref{fig:figure4}(a).  The magnetoresistance measurements in the TSS regime were repeated with different dc currents superposed on the measurement ac current of 10~nA - the data is plotted in figure~\ref{fig:figure4}(c). The critical current extracted from these measurements was about 2~$\mu$A which matches well with the critical current measured in similar systems~\cite{dikin2011coexistence}. To the best of our knowledge, a magnetic field assisted transient superconducting state has not been observed so far. In a related work a slight reduction in resistance on the insulating side of the superconductor-insulator transition was seen whose magnitude depended on $dB/dt$ which was interpreted as a signature of the presence of localized cooper pairs in the system in the non-superconducting state~\cite{mehta2012evidence}. The appearance of the TSS depends on the value of the highest magnetic field $B_{max}$ to which the system is taken before the field is ramped down. We observed that for $B_{max}<6$ T the system does not attain the TSS (see Figure~\ref{fig:figure4}(b)). Interestingly, we also do not observe the TSS when the magnetic field is applied parallel to the interface (see figure~\ref{fig:figureS10}).

There is a due concern about the possible changes in temperature of the sample from changes in spin-entropy due to rapid cycling of the magnetic field. We have calculated this change to be in the order of $\mu$K owing to the low carrier density of the device (Appendix A). We have also checked for measurement artifacts arising due to any remnant field from the superconducting magnet and have ruled them out through careful measurements. Possible effects arising from magnetocaloric effects of the entire sample holder was ruled out by measuring the temperature changes of a bare calibrated temperature sensor of similar thermal mass as the \LAOSTO devies. The sensor was mounted in the chip carrier identically as the \LAOSTO devies - the change in temperature of the sensor due to rapid cycling of the magnetic field was negligibly small.

As a further check we have also performed similar experiments on \STO made conducting by Ar$^+$ ion irradiation~\cite{PhysRevB.91.205117} and on LaTiO$_3$/SrTiO$_3$ heterostructures~\cite{ADMA:ADMA201001980}. Both these systems are known to have a low temperature superconducting behaviour very similar to that of \LAOSTO but lack the competing ferromagnetic order ~\cite{schooley1964superconductivity,biscaras2012two,biscaras2010two}. The measurements on these two systems were performed for exactly the same sample dimensions, gate voltage range, temperature and magnetic field sweep rates as was used for LaAlO$_3$/SrTiO$_3$;  the magnetoresistance data are plotted in figure~\ref{fig:figure7}(a) for reduced \STO and in figure~\ref{fig:figure7}(b) for LaTiO$_3$/SrTiO$_3$ heterostructure.  We find in both cases that the magnetoresistance plots for increasing   and for decreasing magnetic fields fall exactly on top of each other - as expected there is no signature of TSS or hysteresis in the magnetoresistance.

\subsection{Number density extracted from Hall measurement}

To understand the origin of TSS it is first necessary to understand the nature of the mobile charge carriers in the system. In Figure \ref{fig:figure5}(a) we plot $n_s$ extracted from the Hall measurement data assuming a single type of charge carrier in the system. We note that for $V_{g}> V_{g}^*$, $n$ appears to decrease with increase in $V_{g}$; simultaneously the Hall voltage $V_{H}$ develops a slight non-linearity with $B$. The charge carriers being electrons in this case, applying a positive gate voltage is expected to enhance the carrier density $n_s$, as can be seen from the plot of resistance vs $V_{g}$ in Figure \ref{fig:figure5}(b). In Figure \ref{fig:figure5}(a) we also plot the estimated carrier density $n_{calc}$ that would be induced in the system by the gate voltage - the estimate takes into account the electric field dependence of the dielectric constant of the \STO substrate~\cite{PhysRevLett.26.851}. We find that $n_{calc}$ and $n_s$ match very well (to within a geometric factor) for $V_{g} < V_{g}^*$.  For values of gate voltage beyond $V_{g}^*$, $n_s$ begins to drop below the expected range showing that the apparent decrease of $n_s$ with increasing gate voltage cannot be accounted for by the electric field dependence of the dielectric constant of the \STO substrate. The fact that $n_s$ seemingly decreases with increase in $V_{g}$ beyond $V_{g}^*$ indicates that the transport in this regime is best described by a multi-band model~\cite{michaeli2012superconducting}. It is known for \LAOSTO heterostructures that at a certain number density, the system undergoes a Lifshitz transition between light and heavy sub-bands having different symmetries~\cite{joshua2012universal}. The additional carriers introduced are believed to occupy a higher mobility $d_{xz}/d_{yz}$ band near the interface and are responsible for the appearance of superconductivity in the system~\cite{PhysRevB.86.201105, michaeli2012superconducting}.


\begin{figure}[!t]
	\begin{center}
		\includegraphics[width=0.48\textwidth]{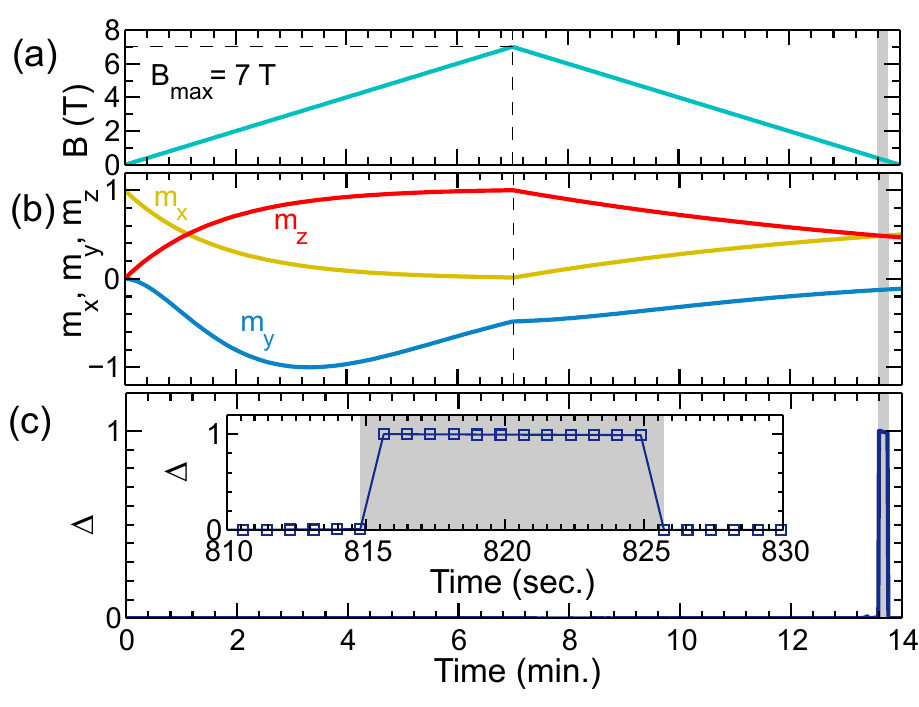}
		\small{{\caption{ (a) Profile of the perpendicular magnetic field varied at the rate  $dB/dt$ = 1 T/min upto a maximum $B_{max}$ = 7~T. (b) Time-variation of  the three components of normalized magnetization, $m_x$, $m_y$ and $m_z$. (c) Time-variation of  the mean-field pairing gap. The shaded regions in all the figures represent the time-window over which superconductivity appears. Inset of figure (c) is the magnified view of the region of non-zero superconducting order parameter. Model parameters used were: hopping amplitude $t' = 0.277$~eV, chemical potential $\mu = 0$, strength of the Rashba spin-orbit interaction $\alpha = 20$~ meV, strength of the attractive pairwise electron-electron interaction $U = t'$, sweep rate $dB/dt = 1$~T/min , maximum applied field $B_{max} = 7$~T, the gyro-magnetic ratio  $\gamma = 1$, $T_{1}$ ($B$ increasing) = 200~sec, $T_{2}$ ($B$ increasing) = 100~sec, $T_{1}$ ($B$ decreasing) = 400 sec, $T_{2}$ ($B$ decreasing) = 200 sec and $T_{3}$ = 550 sec ($T_{1}$, $T_{2}$ and $T_{3}$ are the relaxation times). \label{fig:figure6}}}}
	\end{center}
\end{figure}

\section{Theory}

There exists now indications, both experimental~\cite{wang2011electronic,bert2011direct,li2011coexistence} and theoretical~\cite{mohanta2014phase,mohanta2014oxygen}, that superconductivity at the interface coexists with (in-plane) magnetization in phase segregated regions~\cite{yu2014polarity}. At low gate voltages our particular device is deep inside the ferromagnetic regime as seen from the large hysteresis in the magnetoresistance.  Beyond a certain critical density the system is in a metastable state - the itinerant electrons in the $d_{xz}/d_{yz}$ orbitals favour a superconducting ground state while the in-plane magnetization ~\cite{banerjee2013ferromagnetic}, which originates from the localized magnetic moments at the interface, opposes superconductivity, suppressing superconducting $T_{c}$.  On the application of a perpendicular magnetic field, magnetization of the (in-plane) FM-aligned domains reduces while the out of plane component of magnetization $m_z$ takes on a finite value. 

To understand quantitatively the origin of the TSS, we have computed the three components of magnetization and the superconducting gap parameter at each instant of time when the perpendicular magnetic field is ramped linearly with time. The data are shown in Figure \ref{fig:figure6}. We start with a situation where the in-plane magnetization (taken along the x-axis) has completely destroyed the superconducting order. While increasing magnetic field, the dynamics of the three components of magnetization is described by the following set of Bloch's equations:

\begin{align*}
\frac{dm_x}{dt} &= {\gamma}B_z(t)m_y - \frac{m_x}{T_2} \notag \\
\frac{dm_y}{dt} &= -{\gamma}B_z(t)m_x -   \frac{m_y}{T_2} \notag \\
\frac{dm_z}{dt} &= -\frac{m_z - m_{zs}}{T_1} \notag
\end{align*}

where $\gamma$  is called the Gyromagnetic ratio, $T_{1}$ and $T_{2}$ are the time-scales for the spin-lattice and spin-spin relaxation respectively. $B_{z}(t)$ is increased at the rate $dB/dt$ so as to reach the final value $B_{max}$ . With the initial conditions $m_{x}(t=0)=m_{x0}$, $m_{y}(t=0)=0$, $m_{z}(t = 0)=0$, the solutions to the above equations are 

\begin{align*}
m_x(t) &= {m_{x0}} cos({\gamma}B_z(t)t)e^{-t/T_2} \notag \\
m_y(t) &= -m_{x0} sin({\gamma}B_z(t)t)e^{-t/T_2} \notag \\
m_z(t) &= m_{zs}[1 - e^{-t/T_1}] \notag 
\end{align*}

Therefore, the in-plane magnetization $m_x$ decreases exponentially from its initial value $m_{xo}$ while the out-of-plane magnetization $m_{z}$ grows up to its saturation value $m_{zs}$. Even though the magnetization $m_{y}$ along y-direction was zero initially, it attains a finite value and oscillates over a large range further degrading the electron pairing. Since perpendicular magnetization is much more detrimental to superconductivity than an in-plane one, it is not possible for the superconductivity to appear in this case.

When the magnetic field is decreased at the rate $dB/dt$ from the value $B_{max}$ at which the final magnetizations are $\{m_{xf} ,m_{yf} ,m_{zf}\}$, the set of equations describing the dynamics are:

\begin{align*}
\frac{dm_x}{dt} &= {\gamma}B_z(t)m_y - \frac{m_x}{T_2} \notag \\
\frac{dm_y}{dt} &= -{\gamma}B_z(t)m_x -   \frac{m_y}{T_2} \notag \\
\frac{dm_z}{dt} &= -\frac{m_z}{T_1} \notag
\end{align*}

The solutions to the above equation, with the initial conditions $m_{x}(t=0)=m_{xf}$, $m_{y}(t=0)=m_{yf}$, $m_{z}(t=0)=m_{zf}$, are

\begin{align*}
m_x(t) &= [{m_{xf}}cos({\gamma}B_z(t)t) + m_{yf} sin({\gamma}B_z(t)t)]e^{-t/T_2} \notag \\
m_y(t) &= [{m_{yf}}cos({\gamma}B_z(t)t) - m_{xf} sin({\gamma}B_z(t)t)]e^{-t/T_2} \notag \\
m_z(t) &= m_{zf}[1 - e^{-t/T_1}] \notag 
\end{align*}

While decreasing magnetic field, the localized moments at the interface start establishing the in-plane magnetization again to its initial value $m_{x0}$ according to 
\begin{align*}
m_{x1}(t)&=m_{x0}[1-e^{-t/T_3}]+m_{xf}e^{-t/T_3}
\end{align*}
which accompanies $m_{x}(t)$ in above equation. 

Therefore when the field is ramped down, $m_z$ starts to decay and the in-plane components of magnetization begins to grow towards its zero-field value. However, a finite relaxation time of $m_x$ implies a finite time for the in-plane magnetization to come back to this value. This creates a narrow time-window when the net magnetization is small enough for the superconducting state to be the lower energy state facilitating the
emergence of this novel TSS. Therefore, at 245 mK, superconductivity is a hidden order~\cite{PhysRevB.92.174531} and is masked by the in-plane magnetization - appearing only when the net magnetization is sufficiently low. The fact that no TSS is seen for magnetic field applied parallel to the interface supports this picture. Our calculations confirm the experimental observation that as the magnetic field is decreased beyond a certain rate, there appears a slice of time where all the three components of magnetization are small enough to make electron pairing energetically favourable thus allowing the superconducting state to manifest. The life-time of this TSS obtained from our calculations is about 12 seconds (for $B_{max}$= 7 T and $dB/dt$ = 1 T/min) which is close to the experimentally observed value of about 10 seconds. The critical maximum magnetic field ($\approx$ 6~T), below which the transient superconductivity does not appear, also comes out naturally from our calculations.

\section{Conclusion}
To conclude, in this paper we report the observation of a novel transient superconducting state which appears when a relaxing normal magnetic field reduces the magnetization of the system to a value such that electron pairing becomes energetically favourable. This shows the inherently metastable nature of the superconducting state competing with a magnetic order. The results may have significant impact in understanding the nature of superconductivity in diverse systems like high T$_c$ superconductors and iron pnictide superconductors where superconductivity manifests as a result of electron doping a parent magnetic compound. The coexistence of superconductivity and magnetic order and their controlled tunability using external field open up a new regime of investigation with potential in device applications.

\section*{Acknowledgments}
AB acknowledges funding from Nanomission, Department of Science and Technology (DST), Govt of India and IISc, Bangalore. AT and NM acknowledge IIT, Kharagpur for support. AD and RCB acknowledge funding from the IFCPAR (Project $\#$ IFCPAR 4704-I) and also extend the acknowledgment to DST for the J C Bose Fellowship of RCB and to CSIR-India for financial support through AQuaRIUS Project. AB acknowledges fruitful discussions with H R Krishnamurthy, Vijay Shenoy and Manish Jain.

\section*{Appendix}

\subsection{Estimate of Adiabatic temperature change due to sweeping of magnetic field}

The change in isothermal magnetic entropy due to magnetic field variation is accompanied by an adiabatic temperature change $\Delta T_{ad}$ given by ~\cite{midya2015large}
\begin{equation}
\Delta T_{ad} = - \mu_0 \int_{0}^{H} \frac{T}{C_p}\bigg( \frac{\partial M}{\partial T} \bigg)_H dH
\end{equation}
where $H$ is the magnetic field, $M$ is the magnetization, $T$ is the temperature and $C_p$ is the zero-field heat capacity.

We have calculated the $\bigg( \frac{\partial M}{\partial T} \bigg)_H$ at a given field $H$ using the values of  magnetization $M(H,T)=M_0 m$  obtained by a mean-field (Curie) calculation for s=1/2 systems, with FM T$_c$ set at 200K and the saturation magnetization 0.3 $\mu_B$ as seen from experiments.

The saturation magnetization ($\sim 0.3 \mu_B$) for \LAOSTO sample is given by $M_0=0.3ng \mu_B J^{\prime}$, where $n\simeq 10^{17}$ m$^{-2}$ is the carrier density, $g=2$, $\mu_B=9.27 \times 10^{-24}$ J/T and $J^{\prime}=1/2$.

\begin{figure}[t]
	\begin{center}
		\includegraphics[width=0.48\textwidth]{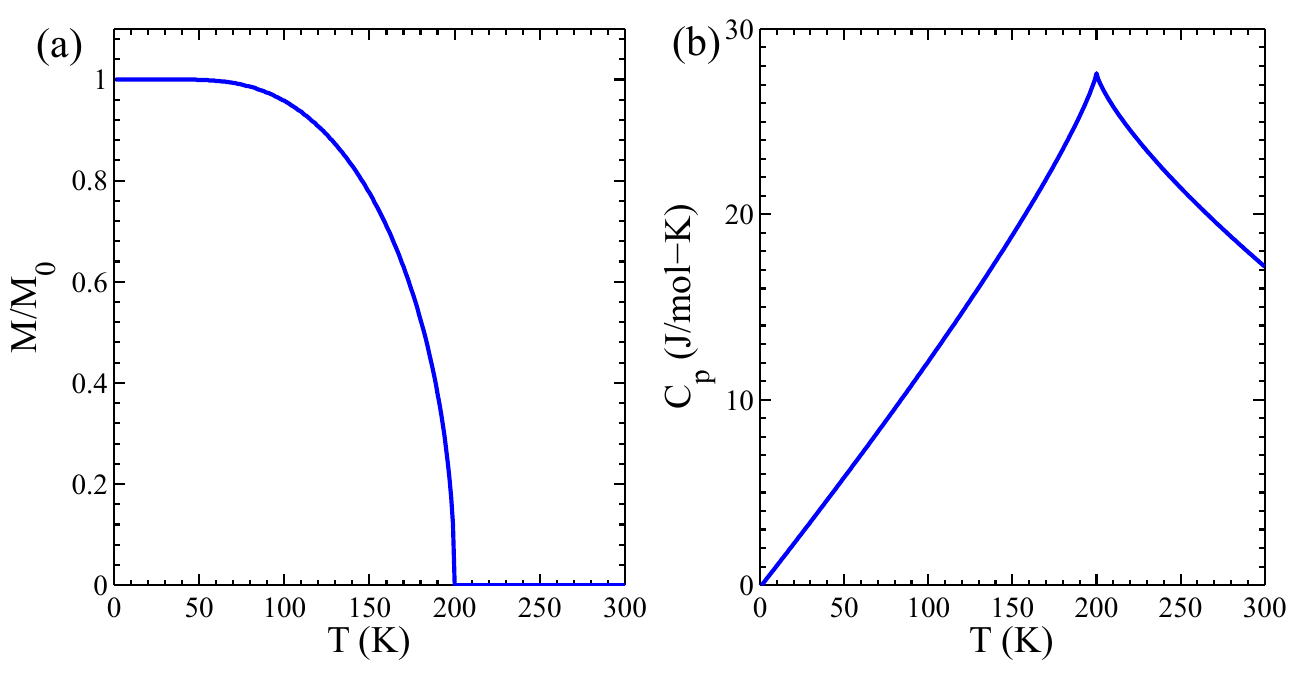}
		\caption{  Temperature variation of the (a) normalized magnetization $M/M_0$ and (b) the specific heat $C_p$.}
		\label{figureS2}\vspace{0em}
	\end{center}
\end{figure}

The heat capacity is given by 
\begin{equation}
C_p=At+(B/\alpha)|t|^{-\alpha}+C
\end{equation}

where $t=T/T_c-1$ and the parameters $A=5$, $B=18$, $\alpha=-0.8$ and $C=27.6$ are used to obtain the temperature dependence of $C_p$ close the typical value for SrTiO$_3$ ~\cite{duran2008specific}. The calculated values of the temperature variations of the normalized magnetization $M/M_0$ and the zero-field specific heat $C_p$ are plotted in figure~\ref{figureS2}.

\begin{figure}[t]
	\begin{center}
		\includegraphics[width=0.48\textwidth]{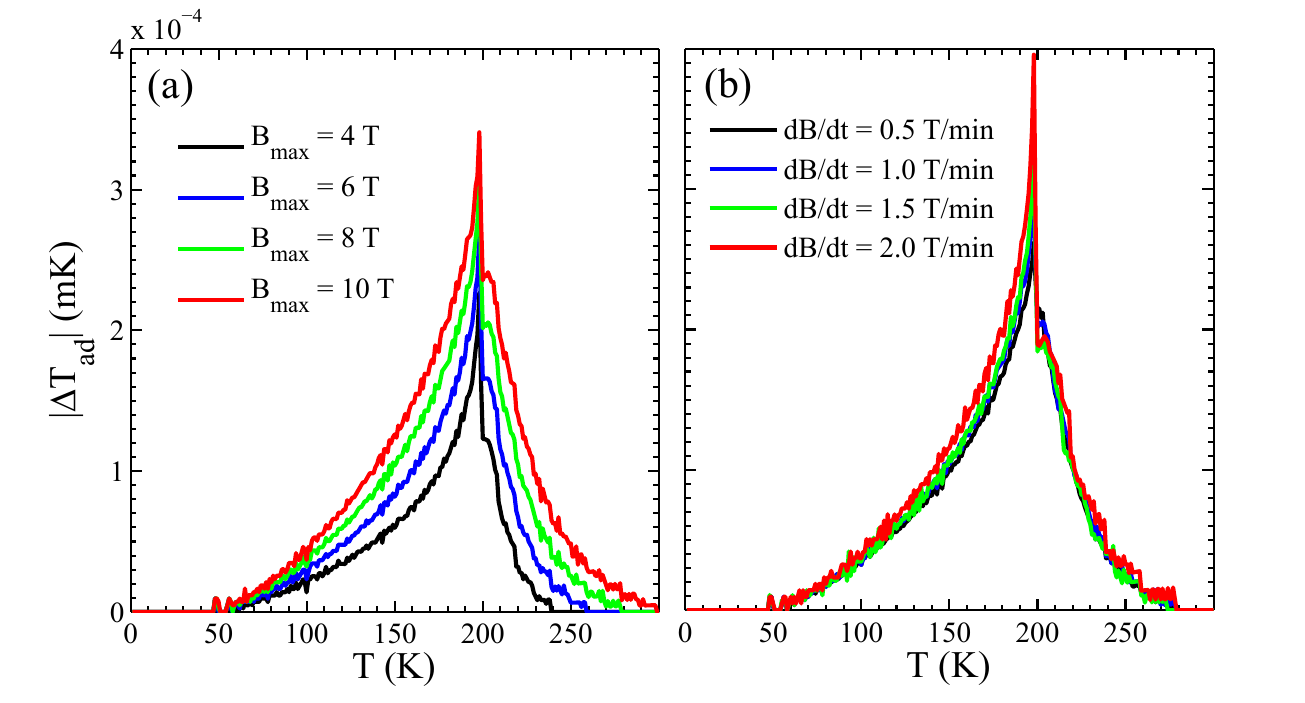}
		\caption{ Temperature variation of the adiabatic temperature change $\Delta T_{ad}$ (a) for different values of the maximum magnetic field $B_{max}$ and constant sweep rate $\frac{dB}{dt}=1$ T/min and (b) for different $\frac{dB}{dt}$ and constant $B_{max}=8$ Tesla.}
		\label{figureS3}\vspace{0em}
	\end{center}
\end{figure}

The resultant adiabatic temperature change is plotted for  different values of maximum magnetic field $B_{max}$ and sweep rate $dB/dt$ in figure~\ref{figureS3}. The nature of the temperature variation is similar to experimental data in other ferromagnetic system~\cite{franco2009field}. The result obtained gives an estimate of the temperature variation due to sweeping of magnetic field and shows that the change in temperature is insignificantly small due to the small carrier density of \LAOSTO interface 2DEG. It shows a peak near the Curie temperature (where maximum entropy is lost) as expected and the temperature variation below 50~K is almost undetectable (see figure~\ref{figureS3}).

The entropy lost due to superconductivity (changes therein due to magnetic field) is exceedingly small since we are deep inside the SC region and the SC $T_c$ actually increases as found experimentally (throwing us deeper in the SC phase).

\begin{figure}[t]
	\begin{center}
		\includegraphics[width=0.48\textwidth]{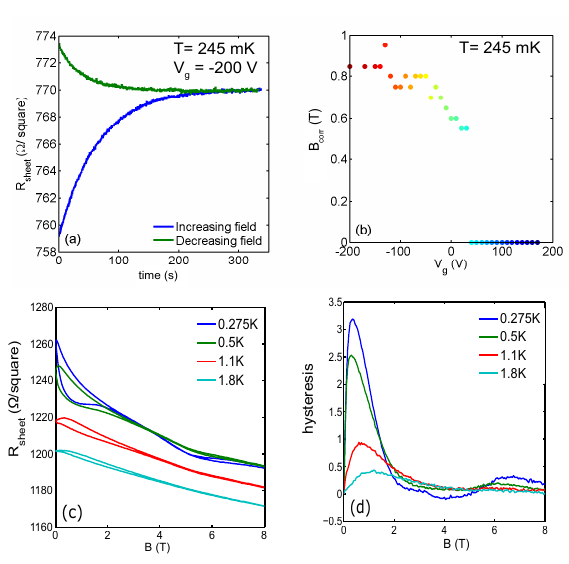}
		\small{\caption{ (a) Relaxation of the magnetoresistance with time.  The magnetic field was initially ramped up from 0 T  to 0.45 T at the rate 1 T/min. The field was then held at 0.45 T and the resistance $R_{sheet}$  monitored as a function of time. In a separate experiment, the magnetic field was ramped down starting from 8 T at the rate 1 T/min to 0.45 T, the field was held at 0.45 T and the resistance monitored as a function of time.  The  measurements were done  at $V_g = -200$ V and temperature 245 mK. (b) Plot as a function of $V_g$ of the magnetic field $B_{corr}$ at which the hysteresis in perpendicular magnetoresistance is maximum. Note that as $V_g$ increases $B_{corr}$ decreases and ultimately vanishes for $V_g > 40$ V. (c) Resistance $R_{sheet}$ as a function of magnetic field at different temperatures at V$_g$ = -200 V. (d) Plot of the hysteresis in magnetoresistance as a function of magnetic field at different temperatures at  V$_g$ = -200 V  - the hysteresis is seen to decrease sharply as the temperature increases. \label{fig:figureS11}}}
	\end{center}
\end{figure}

\subsection{Possible effect of remnant field of the superconducting magnet}
We have checked very carefully for the effect of remnant field from the superconducting magnets and have ruled out its effect on the phase diagram and on the observed transient superconducting state by the following arguments:
\begin{itemize}
	\item{The cryogen-free dilution refrigerator and the magnet in it are frequently warmed up to room temperature. The measurements of Tc reported here have been performed after such a warm up before the magnetic field was turned on.}
	\item{Any effect of remnant magnetic field would result in an asymmetric magnetoresistance scan with the peak of magnetoresistance shifted away from zero magnetic field. As shown in the data in figure \ref{fig:figure2} (a), the magnetoresistance data always peaks at the zero magnetic field showing that the effect of trapped fluxes in the superconducting magnet coil is negligibly small.  }
	\item{The measurements reported here have been performed on two different cryostats  - a cryogen-free dilution refrigerator equipped with a 16 T magnet which operates down to 10 mK and a wet He-3 system equipped with an 8 Tesla magnet which operates down to 250 mK.  The magnets in these two systems are very different in size, inductance and construction. The data obtained in both these cryostats could be compared down to 245 mK (the base temperature of the He-3 system) - till this temperature the data obtained from both superpose on each other showing that the effect of trapped field, if any, is negligibly small.}
\end{itemize}

We have also considered the effects of $dB/dt$ on the copper sample holder. There will be some currents induced due to Faraday Effect but a simple estimate showed these to be negligibly small.

\subsection{Model Hamiltonian and Bogoliubov-de Gennes (BdG) treatment:} We consider the following tight-binding Hamiltonian, to describe electron-pairing at the interface
\begin{eqnarray}
H &=& -t' {\sum_{ \langle ij \rangle ,\sigma}}(c_{i{\sigma}}^{\dagger} c_{j{\sigma}} + H.c.) 
- {\mu} \sum_{i,\sigma}c_{i{\sigma}}^{\dagger} c_{i{\sigma}} 
\notag \\
&& 
- {\mu_B}\sum_{i,\sigma,{\sigma}'}(\mathbf{h} \cdot \mathbf{\sigma})_{\sigma{\sigma}'}c_{i{\sigma}}^{\dagger}c_{i{\sigma}'} \notag\\
&& -i\frac{\alpha}{2} \sum_{\langle ij \rangle, \sigma,{\sigma}'} c_{i{\sigma}}^{\dagger}({\bar{\sigma}}{_{{\sigma}{\sigma}'}}{\times}\bar{d}_{ij})^{z} c_{j{\sigma}'} \notag \\
&&\qquad+ \sum_{i} (\Delta_ic_{i\uparrow}^{\dagger}c_{i\downarrow}^{\dagger} + H.c.)
\end{eqnarray}
\noindent where $t^{'}$ is the kinetic hopping amplitude of electrons, $\mu$ is the chemical potential, $\mu_B$ is the Bohr magneton, $h=(m_{x}, m_{y}, m_{z)}$ represents the exchange fields due to the different components of magnetization, $\alpha$ is the strength of the Rashba spin-orbit interaction, $\bar{d}_{ij}$ is a unit vector between sites $i$ and $j$, and $\Delta_{i}=-U\langle c_{i\uparrow}c_{i\downarrow} \rangle$ is the onsite pairing amplitude with the attractive pair-potential $U$. 

The above Hamiltonian is diagonalized via a spin-generalized Bogoliobov-Valatin transformation 
$\hat{c}_{i\sigma}(\vec{r}_i) = \sum_{i,\sigma'}u_{n{\sigma}{\sigma}'} (\vec{r}_i)\hat{\gamma}_{n{\sigma}'} + 
{v^*}_{n{\sigma}{\sigma}'}(\vec{r}_i)\hat{\gamma^{\dagger}}_{n{\sigma}'}$
and the quasi-particle amplitudes $u_{n\sigma}(\vec{r}_i)$ and $v_{n\sigma}(\vec{r}_i)$  are determined by solving the BdG equations 
\begin{eqnarray}
H{\phi_n}(\vec{r}_i) = {\epsilon_n}{\phi_n}(\vec{r}_i),
\end{eqnarray}
where, ${\phi_n}(\vec{r}_i)$ = $[u_{n,\uparrow}(\vec{r}_i),u_{n,\downarrow}(\vec{r}_i),v_{n,\uparrow}(\vec{r}_i),v_{n,\downarrow}(\vec{r}_i)]$. The local pairing gap $\Delta_i$  is obtained using the following relation:
\begin{eqnarray}
\Delta_{i} = -U{\sum_{n}}[u_{n,\uparrow}(\vec{r}_i){v^*}_{n,\downarrow}(\vec{r}_i)(1 - f(\epsilon_n)) + \notag\\ u_{n,\downarrow}(\vec{r}_i){v^*}_{n,\uparrow}(\vec{r}_i)f(\epsilon_n)] 
\end{eqnarray}
where, $f(x)= 1/(1+exp(x/k_{B}T))$ is the Fermi-Dirac distribution function at temperature $T$ with k$_{B}$, the Boltzmann constant. At any instant of time $t$, the components of magnetization are calculated and then inserted into the above Hamiltonian to solve the mean-field pairing gap self-consistently.

\subsection{Relaxation of the magnetoresistance in perpendicular field} Figure~\ref{fig:figureS11} (a) shows a plot of the relaxation of the resistance $R_{sheet}$ as a function of time at 0.45 T magnetic field. To obtain this data, the magnetic field was initially ramped up from 0~T  to 0.45 T at the rate 1 T/min. The magnet was then held constant at 0.45 T and the resistance monitored as a function of time. It was seen that $R_{sheet}$ relaxes to a higher value over a couple of minutes. In a separate experiment, the magnetic field was ramped down starting from 8 T at the rate 1 T/min to 0.45 T, the field was held at 0.45 T and the resistance monitored as a function of time.  It can be seen that in both cases $R_{sheet}$ relaxes to the same value, although with slightly different time constants.  Figure~\ref{fig:figureS11} (d) shows the temperature evolution of the hysteresis in magnetoresistance. The hysteresis in magnetoresistance weakens as the temperature is increased and eventually dies out by 1.8~K.


\begin{thebibliography}{44}%
	\makeatletter
	\providecommand \@ifxundefined [1]{%
		\@ifx{#1\undefined}
	}%
	\providecommand \@ifnum [1]{%
		\ifnum #1\expandafter \@firstoftwo
		\else \expandafter \@secondoftwo
		\fi
	}%
	\providecommand \@ifx [1]{%
		\ifx #1\expandafter \@firstoftwo
		\else \expandafter \@secondoftwo
		\fi
	}%
	\providecommand \natexlab [1]{#1}%
	\providecommand \enquote  [1]{``#1''}%
	\providecommand \bibnamefont  [1]{#1}%
	\providecommand \bibfnamefont [1]{#1}%
	\providecommand \citenamefont [1]{#1}%
	\providecommand \href@noop [0]{\@secondoftwo}%
	\providecommand \href [0]{\begingroup \@sanitize@url \@href}%
	\providecommand \@href[1]{\@@startlink{#1}\@@href}%
	\providecommand \@@href[1]{\endgroup#1\@@endlink}%
	\providecommand \@sanitize@url [0]{\catcode `\\12\catcode `\$12\catcode
		`\&12\catcode `\#12\catcode `\^12\catcode `\_12\catcode `\%12\relax}%
	\providecommand \@@startlink[1]{}%
	\providecommand \@@endlink[0]{}%
	\providecommand \url  [0]{\begingroup\@sanitize@url \@url }%
	\providecommand \@url [1]{\endgroup\@href {#1}{\urlprefix }}%
	\providecommand \urlprefix  [0]{URL }%
	\providecommand \Eprint [0]{\href }%
	\providecommand \doibase [0]{http://dx.doi.org/}%
	\providecommand \selectlanguage [0]{\@gobble}%
	\providecommand \bibinfo  [0]{\@secondoftwo}%
	\providecommand \bibfield  [0]{\@secondoftwo}%
	\providecommand \translation [1]{[#1]}%
	\providecommand \BibitemOpen [0]{}%
	\providecommand \bibitemStop [0]{}%
	\providecommand \bibitemNoStop [0]{.\EOS\space}%
	\providecommand \EOS [0]{\spacefactor3000\relax}%
	\providecommand \BibitemShut  [1]{\csname bibitem#1\endcsname}%
	\let\auto@bib@innerbib\@empty
	\bibitem [{\citenamefont {Ohtomo}\ and\ \citenamefont
		{Hwang}(2004)}]{ohtomo2004high}%
	\BibitemOpen
	\bibfield  {author} {\bibinfo {author} {\bibfnamefont {A.}~\bibnamefont
			{Ohtomo}}\ and\ \bibinfo {author} {\bibfnamefont {H.}~\bibnamefont {Hwang}},\
	}\href@noop {} {\bibfield  {journal} {\bibinfo  {journal} {Nature}\ }\textbf
	{\bibinfo {volume} {427}},\ \bibinfo {pages} {423} (\bibinfo {year}
	{2004})}\BibitemShut {NoStop}%
\bibitem [{\citenamefont {Sulpizio}\ \emph {et~al.}(2014)\citenamefont
	{Sulpizio}, \citenamefont {Ilani}, \citenamefont {Irvin},\ and\ \citenamefont
	{Levy}}]{doi:10.1146/annurev-matsci-070813-113437}%
\BibitemOpen
\bibfield  {author} {\bibinfo {author} {\bibfnamefont {J.~A.}\ \bibnamefont
		{Sulpizio}}, \bibinfo {author} {\bibfnamefont {S.}~\bibnamefont {Ilani}},
	\bibinfo {author} {\bibfnamefont {P.}~\bibnamefont {Irvin}}, \ and\ \bibinfo
	{author} {\bibfnamefont {J.}~\bibnamefont {Levy}},\ }\href {\doibase
	10.1146/annurev-matsci-070813-113437} {\bibfield  {journal} {\bibinfo
		{journal} {Annual Review of Materials Research}\ }\textbf {\bibinfo {volume}
		{44}},\ \bibinfo {pages} {117} (\bibinfo {year} {2014})}\BibitemShut
{NoStop}%
\bibitem [{\citenamefont {Hwang}\ \emph {et~al.}(2012)\citenamefont {Hwang},
	\citenamefont {Iwasa}, \citenamefont {Kawasaki}, \citenamefont {Keimer},
	\citenamefont {Nagaosa},\ and\ \citenamefont {Tokura}}]{hwang2012emergent}%
\BibitemOpen
\bibfield  {author} {\bibinfo {author} {\bibfnamefont {H.}~\bibnamefont
		{Hwang}}, \bibinfo {author} {\bibfnamefont {Y.}~\bibnamefont {Iwasa}},
	\bibinfo {author} {\bibfnamefont {M.}~\bibnamefont {Kawasaki}}, \bibinfo
	{author} {\bibfnamefont {B.}~\bibnamefont {Keimer}}, \bibinfo {author}
	{\bibfnamefont {N.}~\bibnamefont {Nagaosa}}, \ and\ \bibinfo {author}
	{\bibfnamefont {Y.}~\bibnamefont {Tokura}},\ }\href@noop {} {\bibfield
	{journal} {\bibinfo  {journal} {Nature materials}\ }\textbf {\bibinfo
		{volume} {11}},\ \bibinfo {pages} {103} (\bibinfo {year} {2012})}\BibitemShut
{NoStop}%
\bibitem [{\citenamefont {Reyren}\ \emph {et~al.}(2007)\citenamefont {Reyren},
	\citenamefont {Thiel}, \citenamefont {Caviglia}, \citenamefont {Kourkoutis},
	\citenamefont {Hammerl}, \citenamefont {Richter}, \citenamefont {Schneider},
	\citenamefont {Kopp}, \citenamefont {R{\"u}etschi}, \citenamefont {Jaccard}
	\emph {et~al.}}]{reyren2007superconducting}%
\BibitemOpen
\bibfield  {author} {\bibinfo {author} {\bibfnamefont {N.}~\bibnamefont
		{Reyren}}, \bibinfo {author} {\bibfnamefont {S.}~\bibnamefont {Thiel}},
	\bibinfo {author} {\bibfnamefont {A.}~\bibnamefont {Caviglia}}, \bibinfo
	{author} {\bibfnamefont {L.~F.}\ \bibnamefont {Kourkoutis}}, \bibinfo
	{author} {\bibfnamefont {G.}~\bibnamefont {Hammerl}}, \bibinfo {author}
	{\bibfnamefont {C.}~\bibnamefont {Richter}}, \bibinfo {author} {\bibfnamefont
		{C.}~\bibnamefont {Schneider}}, \bibinfo {author} {\bibfnamefont
		{T.}~\bibnamefont {Kopp}}, \bibinfo {author} {\bibfnamefont {A.-S.}\
		\bibnamefont {R{\"u}etschi}}, \bibinfo {author} {\bibfnamefont
		{D.}~\bibnamefont {Jaccard}},  \emph {et~al.},\ }\href@noop {} {\bibfield
	{journal} {\bibinfo  {journal} {Science}\ }\textbf {\bibinfo {volume}
		{317}},\ \bibinfo {pages} {1196} (\bibinfo {year} {2007})}\BibitemShut
{NoStop}%
\bibitem [{\citenamefont {Mannhart}\ \emph {et~al.}(2008)\citenamefont
	{Mannhart}, \citenamefont {Blank}, \citenamefont {Hwang}, \citenamefont
	{Millis},\ and\ \citenamefont {Triscone}}]{mannhart2008two}%
\BibitemOpen
\bibfield  {author} {\bibinfo {author} {\bibfnamefont {J.}~\bibnamefont
		{Mannhart}}, \bibinfo {author} {\bibfnamefont {D.}~\bibnamefont {Blank}},
	\bibinfo {author} {\bibfnamefont {H.}~\bibnamefont {Hwang}}, \bibinfo
	{author} {\bibfnamefont {A.}~\bibnamefont {Millis}}, \ and\ \bibinfo {author}
	{\bibfnamefont {J.-M.}\ \bibnamefont {Triscone}},\ }\href@noop {} {\bibfield
	{journal} {\bibinfo  {journal} {MRS bulletin}\ }\textbf {\bibinfo {volume}
		{33}},\ \bibinfo {pages} {1027} (\bibinfo {year} {2008})}\BibitemShut
{NoStop}%
\bibitem [{\citenamefont {Zubko}\ \emph {et~al.}(2011)\citenamefont {Zubko},
	\citenamefont {Gariglio}, \citenamefont {Gabay}, \citenamefont {Ghosez},\
	and\ \citenamefont {Triscone}}]{zubko2011interface}%
\BibitemOpen
\bibfield  {author} {\bibinfo {author} {\bibfnamefont {P.}~\bibnamefont
		{Zubko}}, \bibinfo {author} {\bibfnamefont {f.}~\bibnamefont {Gariglio}},
	\bibinfo {author} {\bibfnamefont {M.}~\bibnamefont {Gabay}}, \bibinfo
	{author} {\bibfnamefont {P.}~\bibnamefont {Ghosez}}, \ and\ \bibinfo {author}
	{\bibfnamefont {J.-M.}\ \bibnamefont {Triscone}},\ }\href@noop {} {\bibfield
	{journal} {\bibinfo  {journal} {Annu. Rev. Condens. Matter Phys.}\ }\textbf
	{\bibinfo {volume} {2}},\ \bibinfo {pages} {141} (\bibinfo {year}
	{2011})}\BibitemShut {NoStop}%
\bibitem [{\citenamefont {Boschker}\ \emph {et~al.}(2015)\citenamefont
	{Boschker}, \citenamefont {Richter}, \citenamefont {Fillis-Tsirakis},
	\citenamefont {Schneider},\ and\ \citenamefont {Mannhart}}]{phonon}%
\BibitemOpen
\bibfield  {author} {\bibinfo {author} {\bibfnamefont {H.}~\bibnamefont
		{Boschker}}, \bibinfo {author} {\bibfnamefont {C.}~\bibnamefont {Richter}},
	\bibinfo {author} {\bibfnamefont {E.}~\bibnamefont {Fillis-Tsirakis}},
	\bibinfo {author} {\bibfnamefont {C.~W.}\ \bibnamefont {Schneider}}, \ and\
	\bibinfo {author} {\bibfnamefont {J.}~\bibnamefont {Mannhart}},\ }\href@noop
{} {\bibfield  {journal} {\bibinfo  {journal} {arXiv preprint
			arXiv:1504.04226}\ } (\bibinfo {year} {2015})}\BibitemShut {NoStop}%
\bibitem [{\citenamefont {Richter}\ \emph {et~al.}(2013)\citenamefont
	{Richter}, \citenamefont {Boschker}, \citenamefont {Dietsche}, \citenamefont
	{Fillis-Tsirakis}, \citenamefont {Jany}, \citenamefont {Loder}, \citenamefont
	{Kourkoutis}, \citenamefont {Muller}, \citenamefont {Kirtley}, \citenamefont
	{Schneider} \emph {et~al.}}]{richter}%
\BibitemOpen
\bibfield  {author} {\bibinfo {author} {\bibfnamefont {C.}~\bibnamefont
		{Richter}}, \bibinfo {author} {\bibfnamefont {H.}~\bibnamefont {Boschker}},
	\bibinfo {author} {\bibfnamefont {W.}~\bibnamefont {Dietsche}}, \bibinfo
	{author} {\bibfnamefont {E.}~\bibnamefont {Fillis-Tsirakis}}, \bibinfo
	{author} {\bibfnamefont {R.}~\bibnamefont {Jany}}, \bibinfo {author}
	{\bibfnamefont {F.}~\bibnamefont {Loder}}, \bibinfo {author} {\bibfnamefont
		{L.}~\bibnamefont {Kourkoutis}}, \bibinfo {author} {\bibfnamefont
		{D.}~\bibnamefont {Muller}}, \bibinfo {author} {\bibfnamefont
		{J.}~\bibnamefont {Kirtley}}, \bibinfo {author} {\bibfnamefont
		{C.}~\bibnamefont {Schneider}},  \emph {et~al.},\ }\href@noop {} {\bibfield
	{journal} {\bibinfo  {journal} {Nature}\ }\textbf {\bibinfo {volume} {502}},\
	\bibinfo {pages} {528} (\bibinfo {year} {2013})}\BibitemShut {NoStop}%
\bibitem [{\citenamefont {Bert}\ \emph {et~al.}(2011)\citenamefont {Bert},
	\citenamefont {Kalisky}, \citenamefont {Bell}, \citenamefont {Kim},
	\citenamefont {Hikita}, \citenamefont {Hwang},\ and\ \citenamefont
	{Moler}}]{bert2011direct}%
\BibitemOpen
\bibfield  {author} {\bibinfo {author} {\bibfnamefont {J.~A.}\ \bibnamefont
		{Bert}}, \bibinfo {author} {\bibfnamefont {B.}~\bibnamefont {Kalisky}},
	\bibinfo {author} {\bibfnamefont {C.}~\bibnamefont {Bell}}, \bibinfo {author}
	{\bibfnamefont {M.}~\bibnamefont {Kim}}, \bibinfo {author} {\bibfnamefont
		{Y.}~\bibnamefont {Hikita}}, \bibinfo {author} {\bibfnamefont {H.~Y.}\
		\bibnamefont {Hwang}}, \ and\ \bibinfo {author} {\bibfnamefont {K.~A.}\
		\bibnamefont {Moler}},\ }\href@noop {} {\bibfield  {journal} {\bibinfo
		{journal} {Nature physics}\ }\textbf {\bibinfo {volume} {7}},\ \bibinfo
	{pages} {767} (\bibinfo {year} {2011})}\BibitemShut {NoStop}%
\bibitem [{\citenamefont {Kirtley}\ \emph {et~al.}(2012)\citenamefont
	{Kirtley}, \citenamefont {Kalisky}, \citenamefont {Bert}, \citenamefont
	{Bell}, \citenamefont {Kim}, \citenamefont {Hikita}, \citenamefont {Hwang},
	\citenamefont {Ngai}, \citenamefont {Segal}, \citenamefont {Walker},
	\citenamefont {Ahn},\ and\ \citenamefont {Moler}}]{PhysRevB.85.224518}%
\BibitemOpen
\bibfield  {author} {\bibinfo {author} {\bibfnamefont {J.~R.}\ \bibnamefont
		{Kirtley}}, \bibinfo {author} {\bibfnamefont {B.}~\bibnamefont {Kalisky}},
	\bibinfo {author} {\bibfnamefont {J.~A.}\ \bibnamefont {Bert}}, \bibinfo
	{author} {\bibfnamefont {C.}~\bibnamefont {Bell}}, \bibinfo {author}
	{\bibfnamefont {M.}~\bibnamefont {Kim}}, \bibinfo {author} {\bibfnamefont
		{Y.}~\bibnamefont {Hikita}}, \bibinfo {author} {\bibfnamefont {H.~Y.}\
		\bibnamefont {Hwang}}, \bibinfo {author} {\bibfnamefont {J.~H.}\ \bibnamefont
		{Ngai}}, \bibinfo {author} {\bibfnamefont {Y.}~\bibnamefont {Segal}},
	\bibinfo {author} {\bibfnamefont {F.~J.}\ \bibnamefont {Walker}}, \bibinfo
	{author} {\bibfnamefont {C.~H.}\ \bibnamefont {Ahn}}, \ and\ \bibinfo
	{author} {\bibfnamefont {K.~A.}\ \bibnamefont {Moler}},\ }\href {\doibase
	10.1103/PhysRevB.85.224518} {\bibfield  {journal} {\bibinfo  {journal} {Phys.
			Rev. B}\ }\textbf {\bibinfo {volume} {85}},\ \bibinfo {pages} {224518}
	(\bibinfo {year} {2012})}\BibitemShut {NoStop}%
\bibitem [{\citenamefont {Li}\ \emph {et~al.}(2011)\citenamefont {Li},
	\citenamefont {Richter}, \citenamefont {Mannhart},\ and\ \citenamefont
	{Ashoori}}]{li2011coexistence}%
\BibitemOpen
\bibfield  {author} {\bibinfo {author} {\bibfnamefont {L.}~\bibnamefont
		{Li}}, \bibinfo {author} {\bibfnamefont {C.}~\bibnamefont {Richter}},
	\bibinfo {author} {\bibfnamefont {J.}~\bibnamefont {Mannhart}}, \ and\
	\bibinfo {author} {\bibfnamefont {R.}~\bibnamefont {Ashoori}},\ }\href@noop
{} {\bibfield  {journal} {\bibinfo  {journal} {Nature Physics}\ }\textbf
	{\bibinfo {volume} {7}},\ \bibinfo {pages} {762} (\bibinfo {year}
	{2011})}\BibitemShut {NoStop}%
\bibitem [{\citenamefont {Michaeli}\ \emph {et~al.}(2012)\citenamefont
	{Michaeli}, \citenamefont {Potter},\ and\ \citenamefont
	{Lee}}]{michaeli2012superconducting}%
\BibitemOpen
\bibfield  {author} {\bibinfo {author} {\bibfnamefont {K.}~\bibnamefont
		{Michaeli}}, \bibinfo {author} {\bibfnamefont {A.~C.}\ \bibnamefont
		{Potter}}, \ and\ \bibinfo {author} {\bibfnamefont {P.~A.}\ \bibnamefont
		{Lee}},\ }\href@noop {} {\bibfield  {journal} {\bibinfo  {journal} {Physical
			review letters}\ }\textbf {\bibinfo {volume} {108}},\ \bibinfo {pages}
	{117003} (\bibinfo {year} {2012})}\BibitemShut {NoStop}%
\bibitem [{\citenamefont {Mohanta}\ and\ \citenamefont
	{Taraphder}(2014{\natexlab{a}})}]{mohanta2014phase}%
\BibitemOpen
\bibfield  {author} {\bibinfo {author} {\bibfnamefont {N.}~\bibnamefont
		{Mohanta}}\ and\ \bibinfo {author} {\bibfnamefont {A.}~\bibnamefont
		{Taraphder}},\ }\href@noop {} {\bibfield  {journal} {\bibinfo  {journal}
		{Journal of Physics: Condensed Matter}\ }\textbf {\bibinfo {volume} {26}},\
	\bibinfo {pages} {025705} (\bibinfo {year} {2014}{\natexlab{a}})}\BibitemShut
{NoStop}%
\bibitem [{\citenamefont {Banerjee}\ \emph {et~al.}(2013)\citenamefont
	{Banerjee}, \citenamefont {Erten},\ and\ \citenamefont
	{Randeria}}]{banerjee2013ferromagnetic}%
\BibitemOpen
\bibfield  {author} {\bibinfo {author} {\bibfnamefont {S.}~\bibnamefont
		{Banerjee}}, \bibinfo {author} {\bibfnamefont {O.}~\bibnamefont {Erten}}, \
	and\ \bibinfo {author} {\bibfnamefont {M.}~\bibnamefont {Randeria}},\
}\href@noop {} {\bibfield  {journal} {\bibinfo  {journal} {Nature Physics}\
}\textbf {\bibinfo {volume} {9}},\ \bibinfo {pages} {626} (\bibinfo {year}
{2013})}\BibitemShut {NoStop}%
\bibitem [{\citenamefont {Zhong}\ \emph {et~al.}(2010)\citenamefont {Zhong},
	\citenamefont {Xu},\ and\ \citenamefont {Kelly}}]{PhysRevB.82.165127}%
\BibitemOpen
\bibfield  {author} {\bibinfo {author} {\bibfnamefont {Z.}~\bibnamefont
		{Zhong}}, \bibinfo {author} {\bibfnamefont {P.~X.}\ \bibnamefont {Xu}}, \
	and\ \bibinfo {author} {\bibfnamefont {P.~J.}\ \bibnamefont {Kelly}},\ }\href
{\doibase 10.1103/PhysRevB.82.165127} {\bibfield  {journal} {\bibinfo
		{journal} {Phys. Rev. B}\ }\textbf {\bibinfo {volume} {82}},\ \bibinfo
	{pages} {165127} (\bibinfo {year} {2010})}\BibitemShut {NoStop}%
\bibitem [{\citenamefont {Coey}\ \emph {et~al.}(2016)\citenamefont {Coey},
	\citenamefont {Venkatesan},\ and\ \citenamefont
	{Stamenov}}]{coey2016surface}%
\BibitemOpen
\bibfield  {author} {\bibinfo {author} {\bibfnamefont {J.}~\bibnamefont
		{Coey}}, \bibinfo {author} {\bibfnamefont {M.}~\bibnamefont {Venkatesan}}, \
	and\ \bibinfo {author} {\bibfnamefont {P.}~\bibnamefont {Stamenov}},\
}\href@noop {} {\bibfield  {journal} {\bibinfo  {journal} {Journal of
		Physics: Condensed Matter}\ }\textbf {\bibinfo {volume} {28}},\ \bibinfo
{pages} {485001} (\bibinfo {year} {2016})}\BibitemShut {NoStop}%
\bibitem [{\citenamefont {Pavlenko}\ \emph {et~al.}(2012)\citenamefont
	{Pavlenko}, \citenamefont {Kopp}, \citenamefont {Tsymbal}, \citenamefont
	{Sawatzky},\ and\ \citenamefont {Mannhart}}]{pavlenko2012magnetic}%
\BibitemOpen
\bibfield  {author} {\bibinfo {author} {\bibfnamefont {N.}~\bibnamefont
		{Pavlenko}}, \bibinfo {author} {\bibfnamefont {T.}~\bibnamefont {Kopp}},
	\bibinfo {author} {\bibfnamefont {E.}~\bibnamefont {Tsymbal}}, \bibinfo
	{author} {\bibfnamefont {G.}~\bibnamefont {Sawatzky}}, \ and\ \bibinfo
	{author} {\bibfnamefont {J.}~\bibnamefont {Mannhart}},\ }\href@noop {}
{\bibfield  {journal} {\bibinfo  {journal} {Physical Review B}\ }\textbf
	{\bibinfo {volume} {85}},\ \bibinfo {pages} {020407} (\bibinfo {year}
	{2012})}\BibitemShut {NoStop}%
\bibitem [{\citenamefont {Khalsa}\ \emph {et~al.}(2013)\citenamefont {Khalsa},
	\citenamefont {Lee},\ and\ \citenamefont {MacDonald}}]{khalsa2013theory}%
\BibitemOpen
\bibfield  {author} {\bibinfo {author} {\bibfnamefont {G.}~\bibnamefont
		{Khalsa}}, \bibinfo {author} {\bibfnamefont {B.}~\bibnamefont {Lee}}, \ and\
	\bibinfo {author} {\bibfnamefont {A.}~\bibnamefont {MacDonald}},\ }\href@noop
{} {\bibfield  {journal} {\bibinfo  {journal} {Physical Review B}\ }\textbf
	{\bibinfo {volume} {88}},\ \bibinfo {pages} {041302} (\bibinfo {year}
	{2013})}\BibitemShut {NoStop}%
\bibitem [{\citenamefont {Salluzzo}\ \emph {et~al.}(2009)\citenamefont
	{Salluzzo}, \citenamefont {Cezar}, \citenamefont {Brookes}, \citenamefont
	{Bisogni}, \citenamefont {De~Luca}, \citenamefont {Richter}, \citenamefont
	{Thiel}, \citenamefont {Mannhart}, \citenamefont {Huijben}, \citenamefont
	{Brinkman}, \citenamefont {Rijnders},\ and\ \citenamefont
	{Ghiringhelli}}]{salluzzo2009orbital}%
\BibitemOpen
\bibfield  {author} {\bibinfo {author} {\bibfnamefont {M.}~\bibnamefont
		{Salluzzo}}, \bibinfo {author} {\bibfnamefont {J.}~\bibnamefont {Cezar}},
	\bibinfo {author} {\bibfnamefont {N.}~\bibnamefont {Brookes}}, \bibinfo
	{author} {\bibfnamefont {V.}~\bibnamefont {Bisogni}}, \bibinfo {author}
	{\bibfnamefont {G.}~\bibnamefont {De~Luca}}, \bibinfo {author} {\bibfnamefont
		{C.}~\bibnamefont {Richter}}, \bibinfo {author} {\bibfnamefont
		{S.}~\bibnamefont {Thiel}}, \bibinfo {author} {\bibfnamefont
		{J.}~\bibnamefont {Mannhart}}, \bibinfo {author} {\bibfnamefont
		{M.}~\bibnamefont {Huijben}}, \bibinfo {author} {\bibfnamefont
		{A.}~\bibnamefont {Brinkman}}, \bibinfo {author} {\bibfnamefont
		{G.}~\bibnamefont {Rijnders}}, \ and\ \bibinfo {author} {\bibfnamefont
		{G.}~\bibnamefont {Ghiringhelli}},\ }\href@noop {} {\bibfield  {journal}
	{\bibinfo  {journal} {Physical review letters}\ }\textbf {\bibinfo {volume}
		{102}},\ \bibinfo {pages} {166804} (\bibinfo {year} {2009})}\BibitemShut
{NoStop}%
\bibitem [{\citenamefont {{Seo}}\ \emph {et~al.}(2009)\citenamefont {{Seo}},
	\citenamefont {{Marton}}, \citenamefont {{Choi}}, \citenamefont {{Hassink}},
	\citenamefont {{Blank}}, \citenamefont {{Hwang}}, \citenamefont {{Noh}},
	\citenamefont {{Egami}},\ and\ \citenamefont {{Lee}}}]{aplmultiple}%
\BibitemOpen
\bibfield  {author} {\bibinfo {author} {\bibfnamefont {S.}~\bibnamefont
		{{Seo}}}, \bibinfo {author} {\bibfnamefont {Z.}~\bibnamefont {{Marton}}},
	\bibinfo {author} {\bibfnamefont {W.}~\bibnamefont {{Choi}}}, \bibinfo
	{author} {\bibfnamefont {G.}~\bibnamefont {{Hassink}}}, \bibinfo {author}
	{\bibfnamefont {D.}~\bibnamefont {{Blank}}}, \bibinfo {author} {\bibfnamefont
		{H.}~\bibnamefont {{Hwang}}}, \bibinfo {author} {\bibfnamefont
		{T.}~\bibnamefont {{Noh}}}, \bibinfo {author} {\bibfnamefont
		{T.}~\bibnamefont {{Egami}}}, \ and\ \bibinfo {author} {\bibfnamefont
		{H.}~\bibnamefont {{Lee}}},\ }\href@noop {} {\bibfield  {journal} {\bibinfo
		{journal} {Applied Physics Letters}\ }\textbf {\bibinfo {volume} {95}},\
	\bibinfo {pages} {082107} (\bibinfo {year} {2009})}\BibitemShut {NoStop}%
\bibitem [{\citenamefont {Biscaras}\ \emph {et~al.}(2012)\citenamefont
	{Biscaras}, \citenamefont {Bergeal}, \citenamefont {Hurand}, \citenamefont
	{Grosset{\^e}te}, \citenamefont {Rastogi}, \citenamefont {Budhani},
	\citenamefont {LeBoeuf}, \citenamefont {Proust},\ and\ \citenamefont
	{Lesueur}}]{biscaras2012two}%
\BibitemOpen
\bibfield  {author} {\bibinfo {author} {\bibfnamefont {J.}~\bibnamefont
		{Biscaras}}, \bibinfo {author} {\bibfnamefont {N.}~\bibnamefont {Bergeal}},
	\bibinfo {author} {\bibfnamefont {S.}~\bibnamefont {Hurand}}, \bibinfo
	{author} {\bibfnamefont {C.}~\bibnamefont {Grosset{\^e}te}}, \bibinfo
	{author} {\bibfnamefont {A.}~\bibnamefont {Rastogi}}, \bibinfo {author}
	{\bibfnamefont {R.}~\bibnamefont {Budhani}}, \bibinfo {author} {\bibfnamefont
		{D.}~\bibnamefont {LeBoeuf}}, \bibinfo {author} {\bibfnamefont
		{C.}~\bibnamefont {Proust}}, \ and\ \bibinfo {author} {\bibfnamefont
		{J.}~\bibnamefont {Lesueur}},\ }\href@noop {} {\bibfield  {journal} {\bibinfo
		{journal} {Physical review letters}\ }\textbf {\bibinfo {volume} {108}},\
	\bibinfo {pages} {247004} (\bibinfo {year} {2012})}\BibitemShut {NoStop}%
\bibitem [{\citenamefont {Biscaras}\ \emph {et~al.}(2010)\citenamefont
	{Biscaras}, \citenamefont {Bergeal}, \citenamefont {Kushwaha}, \citenamefont
	{Wolf}, \citenamefont {Rastogi}, \citenamefont {Budhani},\ and\ \citenamefont
	{Lesueur}}]{biscaras2010two}%
\BibitemOpen
\bibfield  {author} {\bibinfo {author} {\bibfnamefont {J.}~\bibnamefont
		{Biscaras}}, \bibinfo {author} {\bibfnamefont {N.}~\bibnamefont {Bergeal}},
	\bibinfo {author} {\bibfnamefont {A.}~\bibnamefont {Kushwaha}}, \bibinfo
	{author} {\bibfnamefont {T.}~\bibnamefont {Wolf}}, \bibinfo {author}
	{\bibfnamefont {A.}~\bibnamefont {Rastogi}}, \bibinfo {author} {\bibfnamefont
		{R.~C.}\ \bibnamefont {Budhani}}, \ and\ \bibinfo {author} {\bibfnamefont
		{J.}~\bibnamefont {Lesueur}},\ }\href@noop {} {\bibfield  {journal} {\bibinfo
		{journal} {Nature communications}\ }\textbf {\bibinfo {volume} {1}},\
	\bibinfo {pages} {89} (\bibinfo {year} {2010})}\BibitemShut {NoStop}%
\bibitem [{\citenamefont {Kim}\ \emph {et~al.}(2010)\citenamefont {Kim},
	\citenamefont {Seo}, \citenamefont {Chisholm}, \citenamefont {Kremer},
	\citenamefont {Habermeier}, \citenamefont {Keimer},\ and\ \citenamefont
	{Lee}}]{kim2010nonlinear}%
\BibitemOpen
\bibfield  {author} {\bibinfo {author} {\bibfnamefont {J.~S.}\ \bibnamefont
		{Kim}}, \bibinfo {author} {\bibfnamefont {S.~S.~A.}\ \bibnamefont {Seo}},
	\bibinfo {author} {\bibfnamefont {M.~F.}\ \bibnamefont {Chisholm}}, \bibinfo
	{author} {\bibfnamefont {R.}~\bibnamefont {Kremer}}, \bibinfo {author}
	{\bibfnamefont {H.-U.}\ \bibnamefont {Habermeier}}, \bibinfo {author}
	{\bibfnamefont {B.}~\bibnamefont {Keimer}}, \ and\ \bibinfo {author}
	{\bibfnamefont {H.~N.}\ \bibnamefont {Lee}},\ }\href@noop {} {\bibfield
	{journal} {\bibinfo  {journal} {Physical Review B}\ }\textbf {\bibinfo
		{volume} {82}},\ \bibinfo {pages} {201407} (\bibinfo {year}
	{2010})}\BibitemShut {NoStop}%
\bibitem [{\citenamefont {Caviglia}\ \emph {et~al.}(2010)\citenamefont
	{Caviglia}, \citenamefont {Gabay}, \citenamefont {Gariglio}, \citenamefont
	{Reyren}, \citenamefont {Cancellieri},\ and\ \citenamefont
	{Triscone}}]{caviglia2010tunable}%
\BibitemOpen
\bibfield  {author} {\bibinfo {author} {\bibfnamefont {A.}~\bibnamefont
		{Caviglia}}, \bibinfo {author} {\bibfnamefont {M.}~\bibnamefont {Gabay}},
	\bibinfo {author} {\bibfnamefont {S.}~\bibnamefont {Gariglio}}, \bibinfo
	{author} {\bibfnamefont {N.}~\bibnamefont {Reyren}}, \bibinfo {author}
	{\bibfnamefont {C.}~\bibnamefont {Cancellieri}}, \ and\ \bibinfo {author}
	{\bibfnamefont {J.-M.}\ \bibnamefont {Triscone}},\ }\href@noop {} {\bibfield
	{journal} {\bibinfo  {journal} {Physical review letters}\ }\textbf {\bibinfo
		{volume} {104}},\ \bibinfo {pages} {126803} (\bibinfo {year}
	{2010})}\BibitemShut {NoStop}%
\bibitem [{\citenamefont {Kalabukhov}\ \emph {et~al.}(2007)\citenamefont
	{Kalabukhov}, \citenamefont {Gunnarsson}, \citenamefont {Borjesson},
	\citenamefont {Olsson}, \citenamefont {Claeson},\ and\ \citenamefont
	{Winkler}}]{kalabukhov2007effect}%
\BibitemOpen
\bibfield  {author} {\bibinfo {author} {\bibfnamefont {A.}~\bibnamefont
		{Kalabukhov}}, \bibinfo {author} {\bibfnamefont {R.}~\bibnamefont
		{Gunnarsson}}, \bibinfo {author} {\bibfnamefont {J.}~\bibnamefont
		{Borjesson}}, \bibinfo {author} {\bibfnamefont {E.}~\bibnamefont {Olsson}},
	\bibinfo {author} {\bibfnamefont {T.}~\bibnamefont {Claeson}}, \ and\
	\bibinfo {author} {\bibfnamefont {D.}~\bibnamefont {Winkler}},\ }\href@noop
{} {\bibfield  {journal} {\bibinfo  {journal} {Physical Review B}\ }\textbf
	{\bibinfo {volume} {75}},\ \bibinfo {pages} {121404} (\bibinfo {year}
	{2007})}\BibitemShut {NoStop}%
\bibitem [{\citenamefont {Ariando}\ \emph {et~al.}(2011)\citenamefont
	{Ariando}, \citenamefont {Wang}, \citenamefont {Baskaran}, \citenamefont
	{Liu}, \citenamefont {Huijben}, \citenamefont {Yi}, \citenamefont {Annadi},
	\citenamefont {Barman}, \citenamefont {Rusydi}, \citenamefont {Dhar},
	\citenamefont {Feng}, \citenamefont {Ding}, \citenamefont {Hilgenkamp},\ and\
	\citenamefont {Venkatesan}}]{wang2011electronic}%
\BibitemOpen
\bibfield  {author} {\bibinfo {author} {\bibnamefont {Ariando}}, \bibinfo
	{author} {\bibfnamefont {X.}~\bibnamefont {Wang}}, \bibinfo {author}
	{\bibfnamefont {G.}~\bibnamefont {Baskaran}}, \bibinfo {author}
	{\bibfnamefont {Z.}~\bibnamefont {Liu}}, \bibinfo {author} {\bibfnamefont
		{J.}~\bibnamefont {Huijben}}, \bibinfo {author} {\bibfnamefont
		{J.}~\bibnamefont {Yi}}, \bibinfo {author} {\bibfnamefont {A.}~\bibnamefont
		{Annadi}}, \bibinfo {author} {\bibfnamefont {A.~R.}\ \bibnamefont {Barman}},
	\bibinfo {author} {\bibfnamefont {A.}~\bibnamefont {Rusydi}}, \bibinfo
	{author} {\bibfnamefont {S.}~\bibnamefont {Dhar}}, \bibinfo {author}
	{\bibfnamefont {Y.}~\bibnamefont {Feng}}, \bibinfo {author} {\bibfnamefont
		{J.}~\bibnamefont {Ding}}, \bibinfo {author} {\bibfnamefont {H.}~\bibnamefont
		{Hilgenkamp}}, \ and\ \bibinfo {author} {\bibfnamefont {T.}~\bibnamefont
		{Venkatesan}},\ }\href@noop {} {\bibfield  {journal} {\bibinfo  {journal}
		{Nature communications}\ }\textbf {\bibinfo {volume} {2}},\ \bibinfo {pages}
	{188} (\bibinfo {year} {2011})}\BibitemShut {NoStop}%
\bibitem [{\citenamefont {Kumar}\ \emph
	{et~al.}(2015{\natexlab{a}})\citenamefont {Kumar}, \citenamefont {Dogra},
	\citenamefont {Bhadauria}, \citenamefont {Gupta}, \citenamefont {Maurya},\
	and\ \citenamefont {Budhani}}]{0953-8984-27-12-125007}%
\BibitemOpen
\bibfield  {author} {\bibinfo {author} {\bibfnamefont {P.}~\bibnamefont
		{Kumar}}, \bibinfo {author} {\bibfnamefont {A.}~\bibnamefont {Dogra}},
	\bibinfo {author} {\bibfnamefont {P.~P.~S.}\ \bibnamefont {Bhadauria}},
	\bibinfo {author} {\bibfnamefont {A.}~\bibnamefont {Gupta}}, \bibinfo
	{author} {\bibfnamefont {K.~K.}\ \bibnamefont {Maurya}}, \ and\ \bibinfo
	{author} {\bibfnamefont {R.~C.}\ \bibnamefont {Budhani}},\ }\href@noop {}
{\bibfield  {journal} {\bibinfo  {journal} {Journal of Physics: Condensed
			Matter}\ }\textbf {\bibinfo {volume} {27}},\ \bibinfo {pages} {125007}
	(\bibinfo {year} {2015}{\natexlab{a}})}\BibitemShut {NoStop}%
\bibitem [{\citenamefont {Kawasaki}\ \emph {et~al.}(1994)\citenamefont
	{Kawasaki}, \citenamefont {Takahashi}, \citenamefont {Maeda}, \citenamefont
	{Tsuchiya}, \citenamefont {Shinohara}, \citenamefont {Ishiyama},
	\citenamefont {Yonezawa}, \citenamefont {Yoshimoto},\ and\ \citenamefont
	{Koinuma}}]{kawasaki1994atomic}%
\BibitemOpen
\bibfield  {author} {\bibinfo {author} {\bibfnamefont {M.}~\bibnamefont
		{Kawasaki}}, \bibinfo {author} {\bibfnamefont {K.}~\bibnamefont {Takahashi}},
	\bibinfo {author} {\bibfnamefont {T.}~\bibnamefont {Maeda}}, \bibinfo
	{author} {\bibfnamefont {R.}~\bibnamefont {Tsuchiya}}, \bibinfo {author}
	{\bibfnamefont {M.}~\bibnamefont {Shinohara}}, \bibinfo {author}
	{\bibfnamefont {O.}~\bibnamefont {Ishiyama}}, \bibinfo {author}
	{\bibfnamefont {T.}~\bibnamefont {Yonezawa}}, \bibinfo {author}
	{\bibfnamefont {M.}~\bibnamefont {Yoshimoto}}, \ and\ \bibinfo {author}
	{\bibfnamefont {H.}~\bibnamefont {Koinuma}},\ }\href@noop {} {\bibfield
	{journal} {\bibinfo  {journal} {Science}\ }\textbf {\bibinfo {volume}
		{266}},\ \bibinfo {pages} {1540} (\bibinfo {year} {1994})}\BibitemShut
{NoStop}%
\bibitem [{\citenamefont {Caviglia}\ \emph {et~al.}(2008)\citenamefont
	{Caviglia}, \citenamefont {Gariglio}, \citenamefont {Reyren}, \citenamefont
	{Jaccard}, \citenamefont {Schneider}, \citenamefont {Gabay}, \citenamefont
	{Thiel}, \citenamefont {Hammerl}, \citenamefont {Mannhart},\ and\
	\citenamefont {Triscone}}]{caviglia2008electric}%
\BibitemOpen
\bibfield  {author} {\bibinfo {author} {\bibfnamefont {A.}~\bibnamefont
		{Caviglia}}, \bibinfo {author} {\bibfnamefont {S.}~\bibnamefont {Gariglio}},
	\bibinfo {author} {\bibfnamefont {N.}~\bibnamefont {Reyren}}, \bibinfo
	{author} {\bibfnamefont {D.}~\bibnamefont {Jaccard}}, \bibinfo {author}
	{\bibfnamefont {T.}~\bibnamefont {Schneider}}, \bibinfo {author}
	{\bibfnamefont {M.}~\bibnamefont {Gabay}}, \bibinfo {author} {\bibfnamefont
		{S.}~\bibnamefont {Thiel}}, \bibinfo {author} {\bibfnamefont
		{G.}~\bibnamefont {Hammerl}}, \bibinfo {author} {\bibfnamefont
		{J.}~\bibnamefont {Mannhart}}, \ and\ \bibinfo {author} {\bibfnamefont
		{J.-M.}\ \bibnamefont {Triscone}},\ }\href@noop {} {\bibfield  {journal}
	{\bibinfo  {journal} {Nature}\ }\textbf {\bibinfo {volume} {456}},\ \bibinfo
	{pages} {624} (\bibinfo {year} {2008})}\BibitemShut {NoStop}%
\bibitem [{\citenamefont {Dikin}\ \emph {et~al.}(2011)\citenamefont {Dikin},
	\citenamefont {Mehta}, \citenamefont {Bark}, \citenamefont {Folkman},
	\citenamefont {Eom},\ and\ \citenamefont
	{Chandrasekhar}}]{dikin2011coexistence}%
\BibitemOpen
\bibfield  {author} {\bibinfo {author} {\bibfnamefont {D.}~\bibnamefont
		{Dikin}}, \bibinfo {author} {\bibfnamefont {M.}~\bibnamefont {Mehta}},
	\bibinfo {author} {\bibfnamefont {C.}~\bibnamefont {Bark}}, \bibinfo {author}
	{\bibfnamefont {C.}~\bibnamefont {Folkman}}, \bibinfo {author} {\bibfnamefont
		{C.}~\bibnamefont {Eom}}, \ and\ \bibinfo {author} {\bibfnamefont
		{V.}~\bibnamefont {Chandrasekhar}},\ }\href@noop {} {\bibfield  {journal}
	{\bibinfo  {journal} {Physical Review Letters}\ }\textbf {\bibinfo {volume}
		{107}},\ \bibinfo {pages} {056802} (\bibinfo {year} {2011})}\BibitemShut
{NoStop}%
\bibitem [{\citenamefont {Brinkman}\ \emph {et~al.}(2007)\citenamefont
	{Brinkman}, \citenamefont {Huijben}, \citenamefont {Van~Zalk}, \citenamefont
	{Huijben}, \citenamefont {Zeitler}, \citenamefont {Maan}, \citenamefont
	{Van~der Wiel}, \citenamefont {Rijnders}, \citenamefont {Blank},\ and\
	\citenamefont {Hilgenkamp}}]{brinkman2007magnetic}%
\BibitemOpen
\bibfield  {author} {\bibinfo {author} {\bibfnamefont {A.}~\bibnamefont
		{Brinkman}}, \bibinfo {author} {\bibfnamefont {M.}~\bibnamefont {Huijben}},
	\bibinfo {author} {\bibfnamefont {M.}~\bibnamefont {Van~Zalk}}, \bibinfo
	{author} {\bibfnamefont {J.}~\bibnamefont {Huijben}}, \bibinfo {author}
	{\bibfnamefont {U.}~\bibnamefont {Zeitler}}, \bibinfo {author} {\bibfnamefont
		{J.}~\bibnamefont {Maan}}, \bibinfo {author} {\bibfnamefont {W.}~\bibnamefont
		{Van~der Wiel}}, \bibinfo {author} {\bibfnamefont {G.}~\bibnamefont
		{Rijnders}}, \bibinfo {author} {\bibfnamefont {D.}~\bibnamefont {Blank}}, \
	and\ \bibinfo {author} {\bibfnamefont {H.}~\bibnamefont {Hilgenkamp}},\
}\href@noop {} {\bibfield  {journal} {\bibinfo  {journal} {Nature materials}\
}\textbf {\bibinfo {volume} {6}},\ \bibinfo {pages} {493} (\bibinfo {year}
{2007})}\BibitemShut {NoStop}%
\bibitem [{\citenamefont {Mehta}\ \emph {et~al.}(2012)\citenamefont {Mehta},
	\citenamefont {Dikin}, \citenamefont {Bark}, \citenamefont {Ryu},
	\citenamefont {Folkman}, \citenamefont {Eom},\ and\ \citenamefont
	{Chandrasekhar}}]{mehta2012evidence}%
\BibitemOpen
\bibfield  {author} {\bibinfo {author} {\bibfnamefont {M.}~\bibnamefont
		{Mehta}}, \bibinfo {author} {\bibfnamefont {D.}~\bibnamefont {Dikin}},
	\bibinfo {author} {\bibfnamefont {C.~W.}\ \bibnamefont {Bark}}, \bibinfo
	{author} {\bibfnamefont {S.}~\bibnamefont {Ryu}}, \bibinfo {author}
	{\bibfnamefont {C.}~\bibnamefont {Folkman}}, \bibinfo {author} {\bibfnamefont
		{C.}~\bibnamefont {Eom}}, \ and\ \bibinfo {author} {\bibfnamefont
		{V.}~\bibnamefont {Chandrasekhar}},\ }\href@noop {} {\bibfield  {journal}
	{\bibinfo  {journal} {Nature Communications}\ }\textbf {\bibinfo {volume}
		{3}},\ \bibinfo {pages} {955} (\bibinfo {year} {2012})}\BibitemShut {NoStop}%
\bibitem [{\citenamefont {Kumar}\ \emph
	{et~al.}(2015{\natexlab{b}})\citenamefont {Kumar}, \citenamefont {Hossain},\
	and\ \citenamefont {Budhani}}]{PhysRevB.91.205117}%
\BibitemOpen
\bibfield  {author} {\bibinfo {author} {\bibfnamefont {D.}~\bibnamefont
		{Kumar}}, \bibinfo {author} {\bibfnamefont {Z.}~\bibnamefont {Hossain}}, \
	and\ \bibinfo {author} {\bibfnamefont {R.~C.}\ \bibnamefont {Budhani}},\
}\href {\doibase 10.1103/PhysRevB.91.205117} {\bibfield  {journal} {\bibinfo
	{journal} {Phys. Rev. B}\ }\textbf {\bibinfo {volume} {91}},\ \bibinfo
{pages} {205117} (\bibinfo {year} {2015}{\natexlab{b}})}\BibitemShut
{NoStop}%
\bibitem [{\citenamefont {Rastogi}\ \emph {et~al.}(2010)\citenamefont
	{Rastogi}, \citenamefont {Kushwaha}, \citenamefont {Shiyani}, \citenamefont
	{Gangawar},\ and\ \citenamefont {Budhani}}]{ADMA:ADMA201001980}%
\BibitemOpen
\bibfield  {author} {\bibinfo {author} {\bibfnamefont {A.}~\bibnamefont
		{Rastogi}}, \bibinfo {author} {\bibfnamefont {A.~K.}\ \bibnamefont
		{Kushwaha}}, \bibinfo {author} {\bibfnamefont {T.}~\bibnamefont {Shiyani}},
	\bibinfo {author} {\bibfnamefont {A.}~\bibnamefont {Gangawar}}, \ and\
	\bibinfo {author} {\bibfnamefont {R.~C.}\ \bibnamefont {Budhani}},\ }\href
{\doibase 10.1002/adma.201001980} {\bibfield  {journal} {\bibinfo  {journal}
		{Advanced Materials}\ }\textbf {\bibinfo {volume} {22}},\ \bibinfo {pages}
	{4448} (\bibinfo {year} {2010})}\BibitemShut {NoStop}%
\bibitem [{\citenamefont {Schooley}\ \emph {et~al.}(1964)\citenamefont
	{Schooley}, \citenamefont {Hosler},\ and\ \citenamefont
	{Cohen}}]{schooley1964superconductivity}%
\BibitemOpen
\bibfield  {author} {\bibinfo {author} {\bibfnamefont {J.}~\bibnamefont
		{Schooley}}, \bibinfo {author} {\bibfnamefont {W.}~\bibnamefont {Hosler}}, \
	and\ \bibinfo {author} {\bibfnamefont {M.~L.}\ \bibnamefont {Cohen}},\ }\href
{\doibase 10.1103/PhysRevLett.12.474} {\bibfield  {journal} {\bibinfo
		{journal} {Phys. Rev. Lett.}\ }\textbf {\bibinfo {volume} {12}},\ \bibinfo
	{pages} {474} (\bibinfo {year} {1964})}\BibitemShut {NoStop}%
\bibitem [{\citenamefont {Sakudo}\ and\ \citenamefont
	{Unoki}(1971)}]{PhysRevLett.26.851}%
\BibitemOpen
\bibfield  {author} {\bibinfo {author} {\bibfnamefont {T.}~\bibnamefont
		{Sakudo}}\ and\ \bibinfo {author} {\bibfnamefont {H.}~\bibnamefont {Unoki}},\
}\href {\doibase 10.1103/PhysRevLett.26.851} {\bibfield  {journal} {\bibinfo
	{journal} {Phys. Rev. Lett.}\ }\textbf {\bibinfo {volume} {26}},\ \bibinfo
{pages} {851} (\bibinfo {year} {1971})}\BibitemShut {NoStop}%
\bibitem [{\citenamefont {Joshua}\ \emph {et~al.}(2012)\citenamefont {Joshua},
	\citenamefont {Pecker}, \citenamefont {Ruhman}, \citenamefont {Altman},\ and\
	\citenamefont {Ilani}}]{joshua2012universal}%
\BibitemOpen
\bibfield  {author} {\bibinfo {author} {\bibfnamefont {A.}~\bibnamefont
		{Joshua}}, \bibinfo {author} {\bibfnamefont {S.}~\bibnamefont {Pecker}},
	\bibinfo {author} {\bibfnamefont {J.}~\bibnamefont {Ruhman}}, \bibinfo
	{author} {\bibfnamefont {E.}~\bibnamefont {Altman}}, \ and\ \bibinfo {author}
	{\bibfnamefont {S.}~\bibnamefont {Ilani}},\ }\href@noop {} {\bibfield
	{journal} {\bibinfo  {journal} {Nature communications}\ }\textbf {\bibinfo
		{volume} {3}},\ \bibinfo {pages} {1129} (\bibinfo {year} {2012})}\BibitemShut
{NoStop}%
\bibitem [{\citenamefont {Fete}\ \emph {et~al.}(2012)\citenamefont {Fete},
	\citenamefont {Gariglio}, \citenamefont {Caviglia}, \citenamefont
	{Triscone},\ and\ \citenamefont {Gabay}}]{PhysRevB.86.201105}%
\BibitemOpen
\bibfield  {author} {\bibinfo {author} {\bibfnamefont {A.}~\bibnamefont
		{Fete}}, \bibinfo {author} {\bibfnamefont {S.}~\bibnamefont {Gariglio}},
	\bibinfo {author} {\bibfnamefont {A.~D.}\ \bibnamefont {Caviglia}}, \bibinfo
	{author} {\bibfnamefont {J.-M.}\ \bibnamefont {Triscone}}, \ and\ \bibinfo
	{author} {\bibfnamefont {M.}~\bibnamefont {Gabay}},\ }\href {\doibase
	10.1103/PhysRevB.86.201105} {\bibfield  {journal} {\bibinfo  {journal} {Phys.
			Rev. B}\ }\textbf {\bibinfo {volume} {86}},\ \bibinfo {pages} {201105}
	(\bibinfo {year} {2012})}\BibitemShut {NoStop}%
\bibitem [{\citenamefont {Mohanta}\ and\ \citenamefont
	{Taraphder}(2014{\natexlab{b}})}]{mohanta2014oxygen}%
\BibitemOpen
\bibfield  {author} {\bibinfo {author} {\bibfnamefont {N.}~\bibnamefont
		{Mohanta}}\ and\ \bibinfo {author} {\bibfnamefont {A.}~\bibnamefont
		{Taraphder}},\ }\href@noop {} {\bibfield  {journal} {\bibinfo  {journal}
		{Journal of Physics: Condensed Matter}\ }\textbf {\bibinfo {volume} {26}},\
	\bibinfo {pages} {215703} (\bibinfo {year} {2014}{\natexlab{b}})}\BibitemShut
{NoStop}%
\bibitem [{\citenamefont {Yu}\ and\ \citenamefont
	{Zunger}(2014)}]{yu2014polarity}%
\BibitemOpen
\bibfield  {author} {\bibinfo {author} {\bibfnamefont {L.}~\bibnamefont
		{Yu}}\ and\ \bibinfo {author} {\bibfnamefont {A.}~\bibnamefont {Zunger}},\
}\href@noop {} {\bibfield  {journal} {\bibinfo  {journal} {Nature
		communications}\ }\textbf {\bibinfo {volume} {5}} (\bibinfo {year}
{2014})}\BibitemShut {NoStop}%
\bibitem [{\citenamefont {Mohanta}\ and\ \citenamefont
	{Taraphder}(2015)}]{PhysRevB.92.174531}%
\BibitemOpen
\bibfield  {author} {\bibinfo {author} {\bibfnamefont {N.}~\bibnamefont
		{Mohanta}}\ and\ \bibinfo {author} {\bibfnamefont {A.}~\bibnamefont
		{Taraphder}},\ }\href {\doibase 10.1103/PhysRevB.92.174531} {\bibfield
	{journal} {\bibinfo  {journal} {Phys. Rev. B}\ }\textbf {\bibinfo {volume}
		{92}},\ \bibinfo {pages} {174531} (\bibinfo {year} {2015})}\BibitemShut
{NoStop}%
\bibitem [{\citenamefont {Midya}\ \emph {et~al.}(2015)\citenamefont {Midya},
	\citenamefont {Mandal}, \citenamefont {Rubi}, \citenamefont {Chen},
	\citenamefont {Wang}, \citenamefont {Mahendiran}, \citenamefont {Lorusso},\
	and\ \citenamefont {Evangelisti}}]{midya2015large}%
\BibitemOpen
\bibfield  {author} {\bibinfo {author} {\bibfnamefont {A.}~\bibnamefont
		{Midya}}, \bibinfo {author} {\bibfnamefont {P.}~\bibnamefont {Mandal}},
	\bibinfo {author} {\bibfnamefont {K.}~\bibnamefont {Rubi}}, \bibinfo {author}
	{\bibfnamefont {R.}~\bibnamefont {Chen}}, \bibinfo {author} {\bibfnamefont
		{J.-S.}\ \bibnamefont {Wang}}, \bibinfo {author} {\bibfnamefont
		{R.}~\bibnamefont {Mahendiran}}, \bibinfo {author} {\bibfnamefont
		{G.}~\bibnamefont {Lorusso}}, \ and\ \bibinfo {author} {\bibfnamefont
		{M.}~\bibnamefont {Evangelisti}},\ }\href@noop {} {\bibfield  {journal}
	{\bibinfo  {journal} {arXiv preprint arXiv:1508.03963}\ } (\bibinfo {year}
	{2015})}\BibitemShut {NoStop}%
\bibitem [{\citenamefont {Dur{\'a}n}\ \emph {et~al.}(2008)\citenamefont
	{Dur{\'a}n}, \citenamefont {Morales}, \citenamefont {Fuentes},\ and\
	\citenamefont {Siqueiros}}]{duran2008specific}%
\BibitemOpen
\bibfield  {author} {\bibinfo {author} {\bibfnamefont {A.}~\bibnamefont
		{Dur{\'a}n}}, \bibinfo {author} {\bibfnamefont {F.}~\bibnamefont {Morales}},
	\bibinfo {author} {\bibfnamefont {L.}~\bibnamefont {Fuentes}}, \ and\
	\bibinfo {author} {\bibfnamefont {J.}~\bibnamefont {Siqueiros}},\ }\href@noop
{} {\bibfield  {journal} {\bibinfo  {journal} {Journal of Physics: Condensed
			Matter}\ }\textbf {\bibinfo {volume} {20}},\ \bibinfo {pages} {085219}
	(\bibinfo {year} {2008})}\BibitemShut {NoStop}%
\bibitem [{\citenamefont {Franco}\ \emph {et~al.}(2009)\citenamefont {Franco},
	\citenamefont {Conde}, \citenamefont {Romero-Enrique}, \citenamefont
	{Spichkin}, \citenamefont {Tishin}, \citenamefont {Zverev} \emph
	{et~al.}}]{franco2009field}%
\BibitemOpen
\bibfield  {author} {\bibinfo {author} {\bibfnamefont {V.}~\bibnamefont
		{Franco}}, \bibinfo {author} {\bibfnamefont {A.}~\bibnamefont {Conde}},
	\bibinfo {author} {\bibfnamefont {J.}~\bibnamefont {Romero-Enrique}},
	\bibinfo {author} {\bibfnamefont {Y.}~\bibnamefont {Spichkin}}, \bibinfo
	{author} {\bibfnamefont {A.}~\bibnamefont {Tishin}}, \bibinfo {author}
	{\bibfnamefont {V.}~\bibnamefont {Zverev}},  \emph {et~al.},\ }\href@noop {}
{\bibfield  {journal} {\bibinfo  {journal} {Journal of Applied Physics}\
	}\textbf {\bibinfo {volume} {106}} (\bibinfo {year} {2009})}\BibitemShut
{NoStop}%
\end{thebibliography}

%

\end{document}